\documentclass[aps,pra,twocolumn,reprint,amsmath,amssymb,graphicx,superscriptaddress]{revtex4-1}
\usepackage{graphicx,amsmath,amssymb,appendix,afterpage,amsfonts,latexsym,color,dcolumn,bm,mathtools}

\newcommand{\beq}{\begin{equation}}
\newcommand{\eeq}{\end{equation}}
\newcommand{\bea}{\begin{eqnarray}}
\newcommand{\eea}{\end{eqnarray}}

\providecommand{\moy}[1]{\langle #1 \rangle}
\providecommand{\bra}[1]{\langle #1 \rvert}
\providecommand{\ket}[1]{\lvert #1 \rangle}

\providecommand{\tr}[1]{\text{tr}\left[ #1 \right]}

\newcommand{\ketbra}[2]{\left| {#1} \right\rangle\left\langle {#2}\right|}
\newcommand{\ud}{\mathrm{d}}

\newcommand{\un}{\openone}

\newcommand{\Leff}{L_{\text{eff}}}

\newcommand{\sP}{\mathcal{P}}
\newcommand{\sQ}{\mathcal{Q}}

\newcommand{\sH}{\mathcal{H}}

\newcommand{\ran}[1]{\text{ran}[#1]}

\graphicspath{ {/} }

\usepackage{tikz}
\usepgflibrary{arrows}
\usetikzlibrary{decorations}
\usetikzlibrary{decorations.pathmorphing,patterns}
\usetikzlibrary{scopes}
\usetikzlibrary{shapes, matrix}
\begin{document}


\title{Adiabatic Elimination and Sub-space Evolution of Open Quantum Systems}

\author{Daniel Finkelstein-Shapiro}
\affiliation{Division of Chemical Physics, Lund University, Box 124, 221 00 Lund, Sweden}
\email{daniel.finkelstein_shapiro@chemphys.lu.se}
\author{David Viennot}
\affiliation{Institut UTINAM,CNRS UMR 6213\\ Universit\'e de Bourgogne-Franche-Comt\'e,
Observatoire de Besan\c{c}on, 25010
Besan\c{c}on, France.}
\author{Ibrahim Saideh}
\affiliation{Laboratoire Mat\'eriaux et Ph\'enom\`enes Quantiques,
Universit\'e Paris Diderot, CNRS UMR 7162, 75013, Paris, France.
Universit\'e Paris-Sud, 91405 Orsay, France}
\author{Thorsten Hansen}
\affiliation{Department of Chemistry, University of Copenhagen, DK 2100 Copenhagen, Denmark}
\email{arne.keller@u-psud.fr}
\author{T\~onu Pullerits}
\affiliation{Division of Chemical Physics, Lund University, Box 124, 221 00 Lund, Sweden}
\author{Arne Keller}
\affiliation{Laboratoire Mat\'eriaux et Ph\'enom\`enes Quantiques,
Universit\'e Paris Diderot, CNRS UMR 7162, 75013, Paris, France.
Universit\'e Paris-Sud, 91405 Orsay, France}
\email{arne.keller@u-psud.fr}

\begin{abstract}
Efficient descriptions of open quantum systems can be obtained by performing an adiabatic elimination of the fast degrees of freedom and formulating effective operators for the slow degrees of freedom in reduced dimensions. 
Here, we perform the construction of effective operators in frequency space, and using the final value theorem or alternatively the Keldysh theorem,  we provide a correction for the trace of the density matrix which takes into account the non trace-preserving character of the evolution. 
We illustrate our results with two different systems, ones where the eliminated fast subspace  is constituted by 
a continuous set of states and ones with discrete states. Furthermore, we show that the two models converge for very large dissipation and at coherent population trapping points. Our results also provide an intuitive picture of the correction to the trace of the density matrix as a detailed balance equation. 
\end{abstract}

\maketitle

\section{Introduction}
The adiabatic elimination method allows to reduce the dimensionality of a problem by discarding fast degrees of freedom and describing only the dynamics of the slow ones.  
Adiabatic elimination  has played an important role in unifying dynamical patterns observed in very different phenomena,  from laser and fluid dynamics to biological and chemical systems~\cite{Haken1975,Haken1977}. 
It has allowed to reduce these apparently very different problems to similar minimal sets of coupled differential equations.
In quantum systems, adiabatic elimination dates back to the sixties in atomic physics, with the development of a theory of the maser and laser which includes the quantum noise due to the spontaneous emission process~\cite{Lax1967}. It has also been essential to understand the mechanisms responsible for atom cooling~\cite{Cohen-tannoudji1992}.
  
While these first applications were concerned with dissipative systems, it seems that in the quantum arena, the adiabatic elimination procedure has been popularized mainly in the case of conservative Hamiltonian systems~\cite{Paulisch_2014,Brion2007,You2003} and, in particular, in many body systems~\cite{Nagy2010,Douglas2015} where it allows one to obtain effective Hamiltonians and open new perspectives for quantum simulations~\cite{Douglas2015}.

Meanwhile, the concept of quantum open systems has emerged and it is now taking over Hamiltonian systems as the elementary brick for the description of a quantum system. 
A quantum open system consists of subsystems interacting with its environment. Its state is described by the density operator, where the degrees of freedom of the bath have been traced out~\cite{Breuer2004}. 
Among quantum open systems, the ones whose dynamics follows a one parameter semigroup play a special role. Indeed, since the work of 
Lindblad, Gorini, Kossakowski and Sudarshan~\cite{Lindblad1976,Gorini1976}, 
the 
form of its generator, the so-called Lindblad operator, is completely 
specified. Furthermore, this specific evolution is the one followed by a 
quantum subsystem interacting with a Markovian environment. 
The concept of open quantum system constitutes a first reduction. Indeed, 
from 
a very high dimensional Hamiltonian dynamics, we end with a Lindblad 
dynamics  
in a Hilbert space of a smaller dimension. But  even this reduced 
description 
can be cumbersome~\cite{Ciuti2018} and to get at least the steady states and 
the dynamics around these steady states can be very difficult and 
computationally intensive. 

When this reduced system Lindblad dynamics presents two different time-scales, it should be useful to separate the fast evolving degrees of freedom from the slow ones, that is, to perform an  adiabatic elimination. In most cases, there is a unique steady state, and the adiabatic elimination consists in obtaining the dynamics in the proximity of the stationary state, where the fast part has already reached a stationary state while the slow part is still evolving to the steady state. In this way, the adiabatic approximation becomes a ``long" time approximation, long with respect to the time needed for the fast part to reach a steady-state behavior. The main objective is then to be able to describe the dynamics of the slow part without the need to refer to the fast one.

To our knowledge the first work which addressed a general formalism to   perform the adiabatic elimination with Lindbladian dynamics is the one by Mirrahimi {\it et al}.~\cite{Mirrahimi2009}. The main idea of this work and subsequent ones~\cite{Azouit2016,Azouit2017,Azouit_structure-preserving_2017,AzouitThesis2017,Forni2018} from the QUANTIC group, consists in preserving the Lindblad structure for the generator of the slow dynamics. To this end they built a bijective map from the exact density matrix to the couple of density matrices corresponding to fast and slow  motions. Using singular perturbation theory~\cite{Tikhonov1952,Fenichel1979,Noethen2011}, they are able, in principle, to obtain the slow motion at any given order of approximation. One of the main points is that
the mapping is such that the dynamics of the slow density matrix is 
generated by an effective Lindblad operator. As a consequence, the dynamics 
of the slow density matrix is trace preserving.

With a completely different methodology, Reiter and S\o rensen obtain an effective Lindblad operator which recovers the same result as obtained in \cite{Mirrahimi2009} (up to an overall energy shift) for the case of a single excited stated, but which can also be applied to more general systems where the energy level structure for the excited states takes into account arbitrary detunings Ref.\cite{Reiter2012}.

We note that in these approaches  the 
density matrix describing the slow part does not accurately  describe the 
quantum state in the slow subspace  
when exchange of population between the fast and slow subspace cannot be 
neglected. Indeed, as the slow dynamics is described by a Lindblad operator, 
it 
is trace preserving and the initial population present in the slow subspace 
will remain in this subspace. 


Adiabatic elimination for many-body systems, in particular for Rydberg 
atoms, has been addressed in~\cite{Lesanovsky2013,Marcuzzi2014} and rely 
mainly in perturbation methods applied to Lindblad operator. In these works, 
the authors calculated the correction up to fourth-order in the perturbation 
and concluded that the physical constraints of the solutions was only 
preserved to second order  
Recently, Macieszczak et al.~\cite{Macieszczak2016} recover a general 
formulation of long time dynamics based on the eigenvalue decomposition of 
the Liouville operator and time dependent perturbation techniques, in order 
to describe a metastable manifold. 
A final application of adiabatic elimination techniques worth noting its 
usefulness in finding conditions for evidence dissipative state preparation 
and noise suppression via interference effects. Recently an extension of 
Ref.~\cite{Reiter2012} presents an effective operator formulation including 
perturbations of the Hamiltonian and of the jump operators involved in 
the dissipative part of the Lindblad operators. They are able to 
show under very general terms how to understand  and implement error 
correction strategies for steady-state subspaces of the Liouvillian 
\cite{Albert2019}. Also, several publications have reported adiabatic 
eliminations in specific 
systems~\cite{Lutkenhaus1998,Damanet2019,Warszawski2000,Burgarth2018}
but without a general recipe to make this approximation.

In this work, we follow an alternate route which consists in using  Feshbach projectors $\sP$ and $\sQ = \un-\sP$~\cite{Feshbach1962} to develop a general strategy to approximate the  evolution of $\sP \rho(t)$, the slow component of the quantum state $\rho(t)$ at time $t$. It is based on the 
the projection $\sP G(z)\sP$  of the resolvent $G(z)= (z-L)^{-1}$ of the original Lindblad operator $L$  in the slow subspace. We define $\Leff(z)$, a $z$--dependent operator defined on the slow subspace only, such that
$\sP G(z)\sP = (z- \Leff(z))^{-1}$. The operator
$L_0=\Leff(z=0)$ is the analog of the effective Lindblad operator  obtained 
previously by Mirrahimi~\cite{Mirrahimi2009} and Reiter~\cite{Reiter2012}. 
Furthermore, we also show how to correct the trace preserving evolution 
generated by $L_0$ to take into account possible population exchange between 
fast and slow subspace.

In this paper, we consider only the case where the  projector $\sP$ onto the space of operators themselves defined on $\sH$ is built from a projector $P$ onto the underlying  Hilbert space $\sH$ as $\sP\rho = P \rho P$, as in Ref.~\cite{Mirrahimi2009, Reiter2012}.
In others words, we assume that the fast/slow partition is linked to a partition of $\sH$ in two complementary subspaces $\sH = P\sH \oplus Q \sH$. The application of our formalism to bipartite systems where the fast/slow partition is linked to a tensorial structure 
$\sH = \sH_{\text{slow}}\otimes \sH_{\text{fast}}$ will be the subject of a future publication.

We apply our general result to several examples where the fast subspace is finite or infinite dimensional. In the last case, we consider that the Hamiltonian of the fast part has a continuous spectrum while the slow part has a discrete one. In other words, we address the problem  of adiabatic elimination of the continuous set of states in dissipative Fano~\cite{Fano1961} systems.

The generalization of Fano interferences from Hamiltonian to open quantum systems whose evolution is generated by a Lindblad operator, has recently been the subject of great interest~\cite{Fano1961,Miroshnichenko2010,Lucky2010,Finkelstein2015,Finkelstein2018,Finkelstein2016-1} in particular to describe mesoscopic systems or condensed matter systems.
In the wide band approximation, corresponding to a ``flat continuum",  we are able to obtain the explicit expression for $\Leff(z)$ and therefore analyze in great detail the adiabatic approximation. 
In particular we show formally and numerically that 
in the limit where the fast dynamics reaches its steady state in a very short time, the Hamiltonian of the fast part can be approximated by a flat continuous spectrum.

The paper is organized as follows: in section II the general formalism is developed and in section III our general results are illustrated with several examples.


\section{Theory}
The Hilbert space $\mathcal{H}$ of the system is partitioned into two subspaces with the help of two orthogonal projectors $P$ and $Q = \un_{\sH} -P$, where $\un_{\sH}$ is the identity operator on $\sH$.
The $Q\mathcal{H}$ subspace represents the fast degrees of freedom which reach a stationary   regime in a short time. 
Our goal is to describe the slow motion in the subspace $P\mathcal{H}$ only, 
after the $Q\mathcal{H}$
has reached its stationary state.

We suppose that the system is coupled to a bath that opens dissipation channels between $Q\sH$ and $P\sH$, or within $P\sH $ and  $Q\sH$. Hamiltonian couplings ($PHQ$, or $QHP$) can also  open transitions between $P\sH$ and $Q\sH$.  
Associated to $P$ and $Q$,  we define super-projector operators $\mathcal{P}$ and $\mathcal{Q}$ such that
\beq
\mathcal{P}\rho = P\rho P; \quad \mathcal{Q} = \un - \mathcal{P},
\eeq
where $\un$ is the identity super-operator on the space of operator on $\mathcal{H}$, and $\rho$ is an operator on $\mathcal{H}$.

We assume that the bath is Markovian so that the density matrix evolves according to a Lindblad's equation \cite{Lindblad1976, Gorini1976}. 
For convenience, we will use the operator-vector 
isomorphism~\cite{Havel2003}, which maps the operator $\ket{a}\bra{b}$ in 
the Hilbert space $\mathcal{H}$ 
onto the vector $\ket{\overline{b}}\otimes\ket{a}$ in the $\mathcal{H}\otimes\mathcal{H}$ Hilbert space, or equivalently maps any $n\times n$ density matrix $\rho$ to a column vector $\vec{\rho}$ with $n^2$ elements, by stacking the columns of the $\rho$ matrix. 
Under this isomorphism, the operation $A\rho B^{\dagger}$ is mapped to 
$\overline{B}\otimes A \vec{\rho}$, where $A$ and $B$ are operators on 
$\mathcal{H}$ and $\overline{B}$ denotes the complex conjugate of $B$; that 
is $\overline{B} = \left(B^{\dagger}\right)^T$, where $B^{\dagger}$ is the 
adjoint and $B^{T}$ is the transpose of $B$ (see 
Appendix~\ref{app:superoper_notation}). 
From now on, we drop the arrow in $\vec{\rho}$ as we assume that $\rho$ is in vector form. The only exception is when a  density matrix $\rho$ is inside a bracket like in $\tr{\rho}$.

With this notation, the super-projectors $\sP$ and $\sQ$ read~:
\beq
\label{eq:defPQ}
\mathcal{P} = P\otimes P; \quad \mathcal{Q} = Q\otimes Q + P \otimes Q + Q \otimes P. 
\eeq
Also, the general form of the Lindblad operator $L$, generator of the 
evolution,  $\dot{\rho} = L\rho$, can be written as\footnote{In the case 
where 
	the system Hamiltonian has a continuous spectrum, the discrete sum can 
	be 
	replaced  by and integral over the generalized Hamiltonian 
	eigenstates}~:
\[
L = -i[\un \otimes H - \bar{H} \otimes \un ] 
+ \sum_i \mathcal{D}(F_i)
\]
where
\beq
\mathcal{D}(F) = 
 \bar{F} \otimes F -\frac{1}{2}(\un \otimes F^{\dagger}F + (F^{\dagger}F)^T 
 \otimes \un) 
 \label{eq:defD}
 \eeq
We start by expressing the density matrix evolution in an integral form 
through the Laplace transform~:
\beq
\label{eq:evolIntegralForm}
\rho(t)=\frac{1}{2\pi i} \int_D e^{zt}G(z)\rho_0 \ud z,
\eeq
where $G(z) = (z-L)^{-1}$ is the resolvent of $L$, and the integral on the complex plane is performed on a straight line $D = \left\{z \in \mathbb{C}; \Re{z}=a>0\right\}$.
Projecting Eq.~\eqref{eq:evolIntegralForm} using $\mathcal{P}$ and $\mathcal{Q}$ gives
\begin{equation}
\begin{split}
\sP\rho(t)&=\frac{1}{2\pi i} \int_D dz e^{zt} (\sP G(z)\sP\rho(0)+\sP G(z)\sQ \rho(0)) \\
\sQ\rho(t)&=\frac{1}{2\pi i} \int_D dz e^{zt} (\sQ G(z)\sP\rho(0)+\sQ G(z)\sQ \rho(0)). \\
\end{split}
\label{eq:resolvents-PQ}
\end{equation}
In the remainder of the text, we make the assumption that at time $t=0$ the population is entirely in the slow subspace $P\mathcal{H}$ so that $\sQ\rho(0)=0$. Hence the evolution in the $P\mathcal{H}$ subspace is simply given by:
\begin{equation}
\sP \rho(t)=\frac{1}{2\pi i} \int \ud z e^{zt} \sP G(z)\sP \rho(0).
\label{eq:P-evolution}
\end{equation}
We define the operator $\Leff(z)$, a $z$-dependent operator defined on $\sP\sH$, such that
$\sP G(z)\sP = \left[z-\Leff(z)\right]^{-1}$.
 Using the definition of the resolvent and the orthogonality of the $\sP$ and $\sQ $ projectors, we have:
\begin{equation}
\label{eq:LeffDef}
\Leff(z) = \sP L \sP+\sP L \sQ G_0(z) \sQ L \sP,
\end{equation} 
where $\sQ G_0(z) \sQ = [z - \sQ L \sQ]^{-1}$ is the resolvent of $\sQ L \sQ$. Equation~\eqref{eq:P-evolution} with Eq.~\eqref{eq:LeffDef} is an exact description of the dynamics (restricted to $P\sH$ subspace) of a system coupled to a Markovian bath, and so is a completely positive map, however it is not trace preserving because the $P\sH$ and $Q\sH$ partitions can exchange population during the evolution. \newline

{\bf Generator of the slow dynamics}. 
We notice that $L_0 = \Leff(z=0)$ is the generator of the slow time dynamics. Indeed, projecting  the Lindblad
equation $\dot{\rho} = L\rho$ on $\sP \sH$ and $\sQ \sH$ we have:
\begin{align}
\mathcal{P}\dot{\rho}(t)&=\mathcal{P}L\mathcal{P}\rho(t)+\mathcal{P}L\mathcal{Q}\rho(t) \label{eq:Pbasic-eq}\\
\sQ\dot{\rho}(t)&=\sQ L\sQ\rho(t)+\sQ L\sP\rho(t)
\label{eq:Qbasic-eq}
\end{align}
To obtain the approximate slow time dynamics in the subspace $P \sH$, we assume that $\sQ \rho$ has reached a stationary regime, $\sQ \dot{\rho}=0$. 
Using Eq.~\eqref{eq:Qbasic-eq} to express $\sQ \rho$ as a function of $\sP \rho$, and inserting the result in Eq.~\eqref{eq:Pbasic-eq}, we obtain:
\begin{equation}
\sQ \dot{\rho} = 0 \Rightarrow \sP \dot{\rho} = L_0 \mathcal{P} \rho.
\label{eq:Leff0}
\end{equation} 
In Appendix~\ref{app:L0TracePreserv}, we show a sufficient condition for $L_0$ to be the generator of a trace preserving evolution.
In all the examples we will present below this condition is fulfilled.  In addition, we have found, explicitly or numerically 
that the operator $L_0$ is of Lindblad form. But we know that we are looking for a non-trace preserving evolution as the total initial population may be distributed on $\sP \rho$ and $\sQ \rho$. We must then correct this evolution to take into account the possible variation of the trace of $\sP \rho$. To this end, we look for the exact final  state, reached in $\sP \sH$ subspace, $\rho_f = \lim_{t\rightarrow\infty}\sP\rho(t)$, from a given initial state $\rho_0 = \rho(t=0)$. \newline

{\bf Mapping to the final state}.  
By Eq.~\eqref{eq:Leff0}, we know that the final state $\rho_f$, in $P\sH$ subspace,  is in the kernel of $L_0 = \Leff(z=0)$. We assume that the kernel is one dimensional and define $\overline{\rho}$ its unique element with $\tr{\overline{\rho}} = 1$. Then $\rho_f= \alpha \overline{\rho}$, and we are let to determine $\alpha = \tr{\rho_f}$. The final stationary state  $\rho_f$ can be obtained taking the limit of Eq.~\eqref{eq:P-evolution} when $t\rightarrow \infty$. This limit can be obtained using the final value theorem:
\beq
\rho_f = \lim_{z\rightarrow 0} z\sP G(z) \sP \rho(0)= 
 \lim_{z\rightarrow 0} z \left[z-\Leff(z)\right]^{-1}\rho(0) .
\eeq
As we show in Appendix~\ref{app:tracecorrection}, this limit can be calculated explicitly as~:
\beq
\rho_f = \alpha\overline{\rho}, \text{ with } \alpha = \frac{1}{\tr{(\un-L_1)\bar{\rho}}}=\frac{1}{1-\langle L_1 \rangle},
\label{eq:finalTrace}
\eeq 
where  $\left. L_1 = \frac{\ud \Leff(z) }{\ud z}\right|_{z=0}$.
We notice that $\alpha$ given by Eq.~\eqref{eq:finalTrace} does not depend on the initial state $\rho_0$. This is a consequence of assuming  that the kernel of $L_0$ is one dimensional. The generalization to the case where the kernel is  multidimensional will be reserved for future work. In this paper we focus on the generic case where the dynamics has only one stationary state. 
The mapping  $\rho_0 \rightarrow \rho_f$ given by Eq.~\eqref{eq:finalTrace} is exact and only requires obtaining $L_1$ and the right eigenvectors of $L_0$. 
Using the definition of $\Leff$, (see Eq.~\eqref{eq:LeffDef}), both operators $L_0 = \Leff(z=0)$ and $L_1 = \left. \frac{\ud \Leff(z) }{\ud z}\right|_{z=0}$ can be written in terms of the original Lindblad operator $L$:
\begin{align}
L_0 &= \sP L \sP - \sP L \sQ \left(\sQ L \sQ\right)^{-1} \sQ L \sP \label{eq:L0def} \\
L_1 &= -\sP L \sQ\left(\sQ L \sQ \right)^{-2} \sQ L \sP \label{eq:L1def}
\end{align}
{\bf Slow time non trace preserving evolution}.
We finally correct the evolution given by Eq.~\eqref{eq:Leff0} by normalizing the state by $\alpha=\tr{\rho_f}$ given by Eq.~\eqref{eq:finalTrace} as:
\beq
\label{eq:correctEvolution}
\rho(t) = \frac{1}{1 - \moy{L_1}} e^{L_0 t}\rho_0.
\eeq
Equation \eqref{eq:correctEvolution} along with  Eqs.~\eqref{eq:L0def} 
and~\eqref{eq:L1def} defining $L_0$ and $L_1$, is one of the main results of 
the paper. 

The difficult part in the calculation of $L_0$ and $L_1$ given by 
Eqs.~\eqref{eq:L0def} 
and~\eqref{eq:L1def} consists in the computation of the inverse of $\sQ L 
\sQ$. As we will see in the next section, this inversion can 
be obtained explicitly only in specific cases. In general, a numerical inversion can be  atempted but can be cumbersome, for instance when  $\ran{Q}$ is an infinite dimensional 
space.   In that case, the inverse can be computed using perturbation theory.
Indeed,  $\sQ L \sQ$ can be written as $\sQ L \sQ = L_D + W$, where the 
matrix 
representation of $L_D$ is diagonal in the basis formed by the eigenvectors 
of $PHP$, $QHQ$ and $W$ is non-diagonal. The inversion of $\sQ L \sQ$ can be 
written as~:
\beq
\label{eq:perturbInversion}
\left(\sQ L \sQ\right)^{-1} = L_D^{-1}\sum_{n=0}^{\infty} 
\left(WL_D^{-1}\right) ^n
\eeq
As we show in appendix~\ref{app:pertubInversion}, in all cases where the 
relaxation processes inside the $\ran{Q}$ subspace can be neglected, $W$ 
will depend only upon the 
Hamiltonian couplings $PHQ$ and $QHP$, and does not depend on the 
dissipative part. The fast dissipation of $Q\sH$ part is involved in $L_D$  
only. Therefore when the adiabatic elimination is a good approximation  it 
is justified to consider that $L_D \gg W$. In most cases, retaining only
the second order terms ($n=2$) at most, in the sum of Eq.~\eqref{eq:perturbInversion}, is enough to obtain a good approximation of 
the dynamics. 
Indeed, the level shift operator ($\Leff (z) - L_0$) of 
Eq.~\eqref{eq:LeffDef} involves the 
operators $\sP L \sQ$ and $\sQ L \sP$ which can be each first or zeroth 
order in the Hamiltonian coupling $QHP$ or $PHQ$, so that only terms 
$n=0,1,2$ for $(\sQ L \sQ)^{-1}$ are 
needed. 

In the next section, we will illustrate in several examples how our
result gives a very good approximation to the true dynamics.

\section{Examples}
We examine the evolution generated by the effective operator $L_0$ derived 
in the previous section with the correction given by 
Eq.~\eqref{eq:correctEvolution}, for a few specific cases when the excited 
states which are eliminated are   i)~continuous manifolds and ii)~discrete 
states. 
We use continuous  manifolds because they are part of fundamental toy models for both basic quantum evolution and spectroscopy, and also because they allow  simplifications in the wide band approximation. 
In such an approximation,  analytical expression of $\Leff(z)$ can be obtained.
 In general, using a continuous set of states in the wide band approximation,  instead of a set of discrete levels,  gives a zero real part of the level-shift operator (also called self-energy) leaving only the imaginary dissipative contribution. 
We then investigate systems with discrete excited states since they are more prevalent. 
We finally show that in the limit of large dissipation the adiabatic evolution where  continuous and discrete excited state manifolds are eliminated  coincide. 
We only consider time independent Hamiltonians, however it can describe the 
case where coherent radiation couples  and excited states but in the 
rotating wave approximation so that all coupling elements are 
time-independent and the detuning between excited and ground states has been 
offset by the energy of the impinging photons. 
 
\subsection{Elimination of continua excited states}

\begin{figure}[ht]
\centering
\begin{tikzpicture}[scale=0.7]
\draw[thick] (0,0) -- (2cm,0) node[below]{$\ket{g_1}$};
\draw[thick] (3,0) -- (5cm,0) node[below]{$\ket{g_2}$};
\draw[thick,dashed] (5.5cm,0) -- (6.5cm,0);
\draw[thick] (7,0) -- (9cm,0) node[below]{$\ket{g_{N_g}}$};
\draw[fill=gray]  (0cm,2cm) rectangle (2cm,5cm) node[right]{$\ket{k_1}$} ;
\draw[fill=gray]  (3cm,2cm) rectangle (5cm,5cm) node[right]{$\ket{k_2}$} ;
\draw[fill=gray]  (7cm,2cm) rectangle (9cm,5cm) node[right]{$\ket{k_{N_e}}$} ;
\draw [<->,thick] (0.3cm,0cm)--(0.3cm,3.cm)
node[midway, left] {$V_1^{(1)}$};
\draw [->,thick,decorate,decoration=snake] (0.6cm,3.cm)--(0.6cm,0) node[pos=0.7,right] {$\Gamma^{(1)}_1$};
\draw [<->,thick] (0.9cm,0cm)--(3.3cm,3.cm)
node[pos=0.55, left] {$V_1^{(2)}$};
\draw [->,thick,decorate,decoration=snake] (3.6cm,3.cm)--(1.2cm,0) node[pos=0.45,right] {$\Gamma^{(2)}_1$};
\draw[<->,thick] (1.5cm,0cm) to[out=45,in=135] node [sloped, above, pos=0.6] {$V_{12}$} (3.2cm,0cm);
\end{tikzpicture}
\caption{\label{fig:FanoDissip} Energy levels and transitions of a Fano-type model with dissipation. Hamiltonian couplings are indicated by straight arrows, dissipative processes by twisted arrows.}
\end{figure}
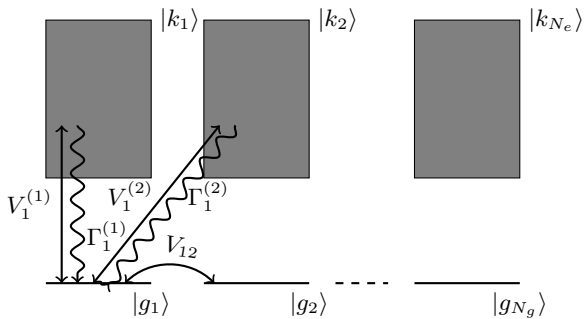

Hamiltonians with continuous spectrum have been part of the spectroscopist toolbox for several decades to describe atomic, molecular and condensed matter systems  \cite{Fano1935, Fano1961,Baldini1962,Jain1965, Glutsch1994,Siegner1995,Siegner1995a,
Siegner1996,Seisyan2016,Holfeld1998,
Yoshino2015,Miroshnichenko2010,Lucky2010}. 
Their distinctive property is that they result in an asymmetric profile arising from interference processes \cite{Fano1961}. 
The Hamiltonian structure as well as dissipative transitions are shown in Figure~\ref{fig:FanoDissip}. 
A set of $N_g$ ground states $\ket{g_i}$ are coupled among themselves by 
Hamiltonian couplings $V_{ij}$ ($i,j = 1, \cdots N_g$) as well as to $N_e$ 
continuous sets of excited states $\ket{k_j}$.  Continua are not coupled  
among themselves (any coupling between continua can be removed by a unitary 
transformation which redefines all the other couplings), they are coupled to 
the ground states 
through Hamiltonian couplings $V_i^{(j)}$  and through dissipation at rates $\Gamma^{(j)}_i$ ($i=1,2,\cdots,N_g$ and $j=1,2,\cdots,N_e$).
In the following we adopt the wide band approximation where the couplings $V_i^{(j)}$, the rates
$\Gamma^{(j)}_i$ and the density of states $n^{(j)} = \frac{\ud k_j}{\ud E}$ per unit of energy $E$, are considered to be independent of $k_j$.

The general problem with $N_g$ ground states coupled to $N_e$ excited states is considered in Appendix~\ref{app:general_continua}, while in the following we examine in detail the case of one continuum coupled to either one or two ground states. \newline

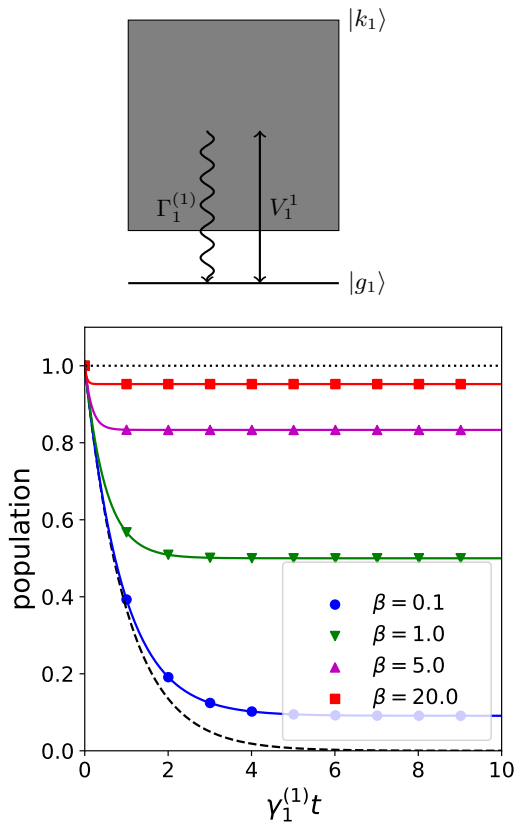
\begin{figure}[h]
\centering
\begin{tikzpicture}[scale=0.7]
\draw[thick] (0,0) -- (4cm,0) node[right]{$\ket{g_1}$};
\draw[fill=gray]  (0cm,1cm) rectangle (4cm,5cm) node[right]{$\ket{k_1}$} ;
\draw[<->,thick] (2.5cm,0) --(2.5cm,2.9cm) node[midway, right] {$V_{1}^{1}$};
\draw [->,thick,decorate,decoration=snake] (1.5cm,2.9cm)--(1.5cm,0)
node[midway,left] {$\Gamma_1^{(1)}$};
\end{tikzpicture}	\includegraphics[width=0.4\textwidth]{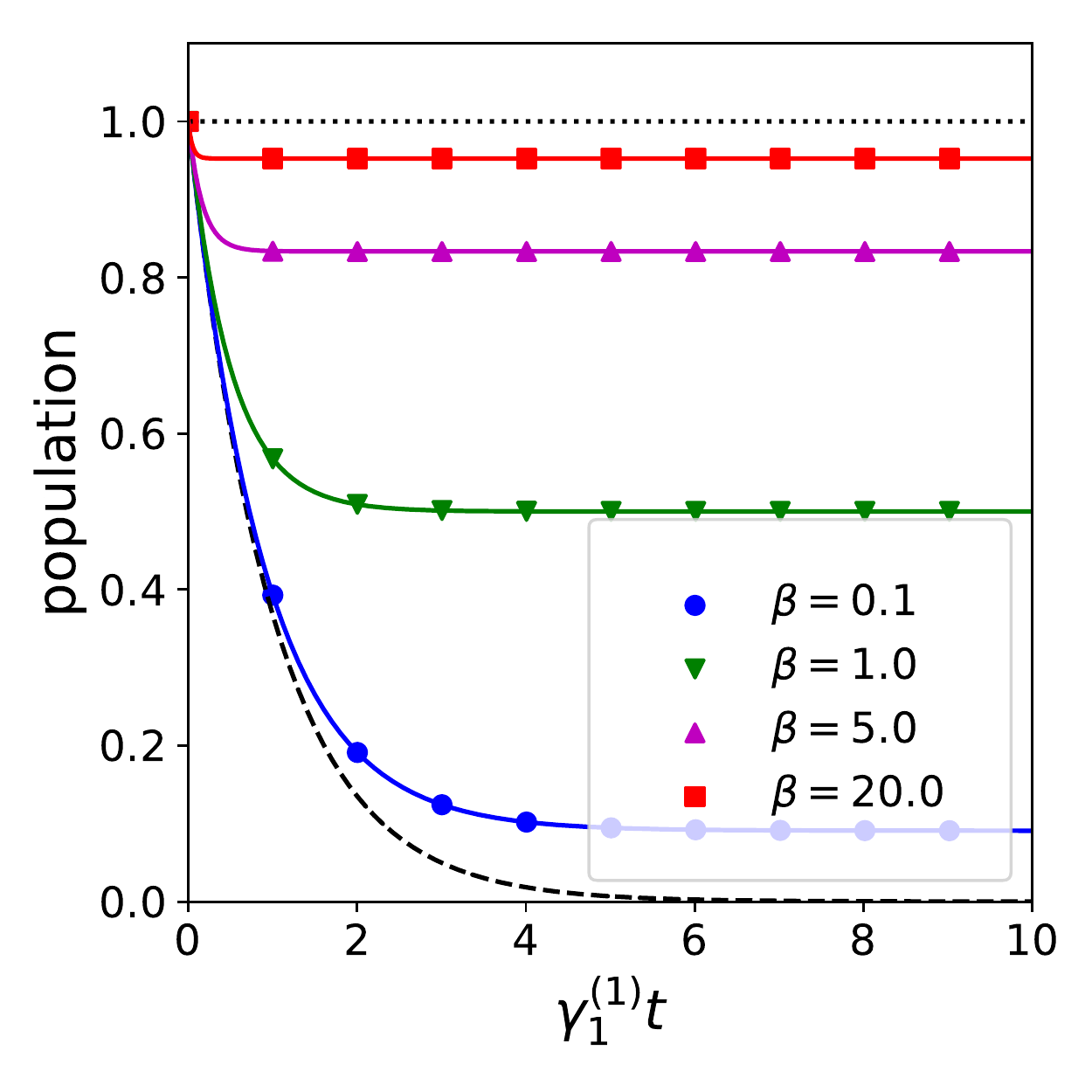}
	\caption{Ground state population of a single discrete system coupled to a continuum as a function of the rescaled time $\gamma_1^{(1)}t$ with $\gamma_1^{(1)}=n\pi(V_1^{(1)})^2$, for different values of the dimensionless constant $\beta=\Gamma_{1}^{(1)}/\gamma_{1}^{(1)}$. The dotted line corresponds to the limit $\Gamma_{1}^{(1)} \to \infty$ which in this case leaves the population unchanged in the discrete state, and the dashed line corresponds to the limit $\Gamma_{1}^{(1)}=0$, which is the limit of a discrete level unitarily coupled to a continuum of states that corresponds to a description for particle decay.}
	\label{fig:continuum_11}
\end{figure}

\textbf{Single ground state level coupled to a single continuum}. 
We first consider a single discrete level coupled to a continuum of states via a Hamiltonian coupling $V_{1}^{(1)}$. The continuum can dissipate back to the ground state wiht a rate $\Gamma_1^{(1)}$ (Fig. \ref{fig:continuum_11}). The Liouvillian for this system is $L = -i(1\otimes H -\bar{H}\otimes 1)+\mathcal{D}(F_1^{(1)} )$ (see Eq.~\eqref{eq:defD}) where:
\begin{equation}
\begin{split}
H &= \int dk_1 V_1^{(1)}\ketbra{g_1}{k_1} + \text{c.c} \\
F_1^{(1)} &= \sqrt{\Gamma_1^{(1)}}\int dk_1 \ketbra{g_1}{k_1}.
\end{split}
\end{equation}
The effective operator $\Leff(z)$ to describe the ground state dynamics after elimination of the continuous set of  excited states can be obtained explicitly using Eq.~\eqref{eq:LeffDef} (see Appendix~\ref{app:general_continua}):
\begin{equation}
\begin{split}
L_{\text{eff}}(z)& = \left(\frac{-z}{z+\Gamma_1^{(1)}} \right) \bar{F}_{\text{eff}} \otimes F_{\text{eff}}
\end{split}
\end{equation}
with $F_{\text{eff}} = \sqrt{\gamma_{1}^{(1)}} \ketbra{g_1}{g_1}$, and where $\gamma_{1}^{(1)}=2n^{(1)}\pi (V_1^{(1)})^2$; it represents the injection rate from discrete to continuum due to the Hamiltonian coupling.  
The operator can be expanded in powers of $z$ as $\Leff(z) \approx z L_1$ where $L_1 = - \frac{\bar{F}_{\text{eff}} \otimes F_{\text{eff}}}{\Gamma_1^{(1)}}$ and where the $z$-independent term $L_0$ is zero.
This means that the approximate dynamics  given by $e^{L_0t} = \un$ (see Eq.~\eqref{eq:correctEvolution}) has no dynamics.
The correction to the ground state is then $\langle 1-L_1 \rangle^{-1}=\frac{1}{1+\beta^{-1}}$ where $\beta=\Gamma_{11}/\gamma_{1}^{(1)}$.
 
 We can readily solve the exact dynamics of the ground state in terms of the dimensionless constant $\beta$ and the rescaled time $\tau=\gamma_{1}^{(1)}t$:
\begin{equation}
\rho(\tau)=\frac{\beta+e^{-(\beta+1)\tau}}{\beta+1}\ketbra{g_1}{g_1}
\label{eq:continuum_11}
\end{equation} 
where we can see that the correction introduced by $\langle 1- L_1 \rangle^{-1}$ is exact. We can already see from this simple example that this correction is nothing else than the detailed balance obtained from a kinetic equation between two sites $\sP$ and $\sQ$ in the steady-state. Indeed, considering temporarily that $\sP$ and $\sQ$ are sites connected by classical rates, and taking $n_{\sP}$ and $n_{\sQ}$ to be the populations of the two sites and $k_{\sQ \to \sP},k_{\sP \to \sQ}$ the transition rates, we can write:
\begin{equation}
\begin{split}
\dot{n}_{\sP} &= k_{\sQ \to \sP} n_{\sQ} - k_{\sP \to \sQ} n_{\sP}\\
\dot{n}_{\sQ} &= -k_{\sQ \to \sP} n_{\sQ} + k_{\sP \to \sQ} n_{\sP}
\end{split}
\end{equation}
which readily yield the steady-state population in $\sP$ as:
 $n_{\sP}=\frac{1}{1 + \frac{k_{Q \to P}}{k_{P \to Q}}}$. We thus identify  $k_{Q \to P}\equiv \Gamma_1^{(1)}$ and $k_{\sP \to \sQ}\equiv \gamma_{1}^{(1)} = n^{(1)}\pi (V_1^{(1)})^2$. The relevant decay from $\sQ$ to $\sP$ is the relaxation rate while the relevant transition from $\sP$ to $\sQ$ is the Hamiltonian rate $\gamma_{1}^{(1)}$. This identification will be recovered in the more complicated case of a two-level system coupled to a continuum and then in a different form in the case of a $\Lambda$ system. 

It is also illustrative to look at the exact solution given by Eq.~\eqref{eq:continuum_11} in the two limits of absent ($\Gamma_1^{(1)}=0$) and very large dissipation ($\Gamma_1^{(1)}\gg \gamma_{1}^{(1)}$) from continuum to the ground state. 
As the dissipation rate $\Gamma_1^{(1)}$ goes to zero, we have a discrete level coupled to a continuum trough Hamiltonian couplings only. This  is the standard model for particle decay or injection into a band~\cite{May2011,Schatz2002}. 
The evolution of the discrete state only,  can be fully described  by a non-Hermitian Hamiltonian alone, entirely in Hilbert space without the need for a Lindblad operator. In this case, the final state has zero population in the discrete ground state as all the population has been lost in the continuum. 
The opposite limit of infinitely high dissipation results in no dynamics whatsoever with the single discrete level being always populated. Because both cases are expressed in superoperator space as the limits of a continuous function of $\Gamma_1^{(1)}$, we provide a rigorous connection between non-Hermitian Hamiltonian decay dynamics ($\Gamma_1^{(1)}/\gamma_{11} \to 0$) and fully trace preserving dissipative dynamics ($\Gamma_1^{(1)}/\gamma_{11} \to \infty$) thanks to the nonlinear term of the form $-\frac{z}{z+\Gamma}$. 
This connection is not restricted to the single discrete level-system but is 
a general feature of discrete levels coupled to a manifold of continua where 
the evolution presents a transition from non-Hermitian decay Hamiltonians to 
trace preserving generators,   when the dissipation rate from the continuum 
is varied, and which could provide insight into comparisons of both 
approaches \cite{Zloshchastiev2014,Echeverri-Arteaga2019}.

\textbf{Two discrete states coupled to a single continuum}. 
The model of a two-level system coupled to a continuous set of states is the 
standard Fano model invoked so often in 
spectroscopy~\cite{Miroshnichenko2010}. Once more, the Liouvillian is 
written as $L = -i(1\otimes H -\bar{H}\otimes 1)+\sum_{i=1,2} 
\mathcal{D}(F_i^{(1)})$ where the Hamiltonian $H = H_0+H_V$ is:
\begin{align}
&H_0=E_{1}\ket{g_1}\bra{g_1}+E_2\ket{g_2}\bra{g_2}+\int dk \epsilon_{k_1}\ket{k_1}\bra{k_1} \nonumber \\
&H_V=V_{12}\ket{g_1}\bra{g_2}+V_{12}^*\ket{g_2}\bra{g_1} \nonumber \\
&+ \int dk_1 \left[V_1^{(1)}\ket{g_1}\bra{k_1}+V_1^{(1)*}\ket{k_1}\bra{g_1}\right] \nonumber \\
&+\int dk_1 
\left[V_2^{(1)} \ket{g_2}\bra{k_1}+V_2^{(1)*}\ket{k_1}\bra{g_2}\right],
\label{eq:Hamiltonian}
\end{align}
and the quantum jump operators are:
\begin{equation}
\begin{split}
F_1^{(1)} = \int dk_1 \sqrt{\Gamma_{1}^{(1)}} \ketbra{g_1}{k_1} \\
F_2^{(1)} = \int dk_1 \sqrt{\Gamma_{2}^{(1)}} \ketbra{g_2}{k_1} \\
\end{split}
\end{equation}
Using Eq.~\eqref{eq:LeffDef} for the effective operator $\Leff(z)$ (see Appendix~\ref{app:general_continua}), we obtain:
\begin{equation}
\Leff(z) = L_0 + \sum_{i=1,2} \Delta_i(z) J_i
\end{equation}
where $L_0 = -i(1\otimes PHP -P\bar{H}P\otimes 1)+\sum_{i=1,2} \mathcal{D}(F_{\text{eff},i})$ and
\begin{equation}
\begin{split}
F_{\text{eff},i} & = \sum_{j=1,2} \sqrt{\frac{\Gamma_{i}^{(1)}}{\Gamma}} V_{j}^{(1)} \ketbra{g_i}{g_j} \\
J_i&= \bar{F}_{\text{eff},i}^{(1)}\otimes F_{\text{eff},i}^{(1)} \\
\Delta_i(z) &= -\frac{z}{(z+\Gamma)} 
\end{split}
\end{equation}
and $\Gamma = \Gamma_{1}^{(1)}+\Gamma_{2}^{(1)}$.
The effective Liouvillian can be expanded in powers of $z$ as:
\begin{equation}
\Leff(z) = L_0 + zL_1 + ...
\end{equation}
where 
\begin{equation}
L_1= -\sum_{i=1,2} \frac{J_i}{\Gamma} 
\end{equation}
and the correction coefficient is
$\alpha = \moy{\un-L_1}^{-1}$ as in Eq.~\eqref{eq:finalTrace}. 
 
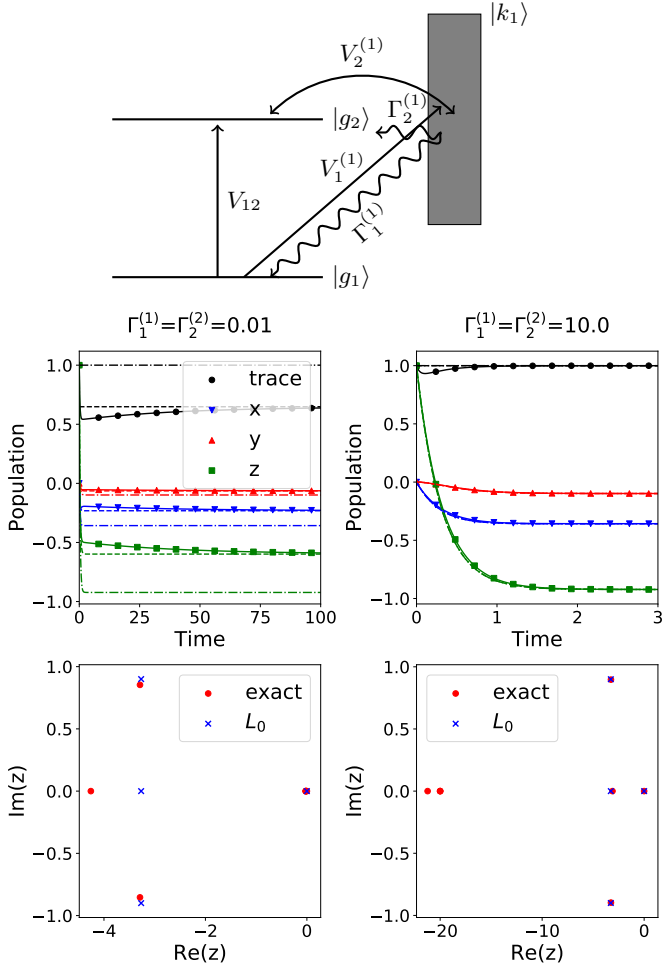
\begin{figure}[h]	
\centering
\begin{tikzpicture}[scale=0.7]
\draw[thick] (0,0) -- (4cm,0) node[right]{$\ket{g_1}$};
\draw[thick] (0,3cm)--(4cm,3cm) node[right]{$\ket{g_2}$};
\draw[fill=gray]  (6cm,1cm) rectangle (7cm,5cm) node[right]{$\ket{k_1}$} ;
\draw[->,thick] (2cm,0) --(2cm,2.9cm) node[midway, right] {$V_{12}$};
\draw [->,thick] (2.5cm,0cm)--(6.25cm,3.25cm)
node[midway, above] {$V_1^{(1)}$};
\draw[<->,thick] (3cm,3.1cm) to[out=45,in=135] node [sloped, above] {$V_2^{(1)}$} (6.5cm,3.1cm);
\draw [->,thick,decorate,decoration=snake] (6.25cm,2.75cm)--(3cm,0)
node[midway,sloped ,below,] {$\Gamma_1^{(1)}$};
\draw [->,thick,decorate,decoration=snake] (6.25cm,2.75cm)--(5cm,2.75)
node[midway,sloped ,above,] {$\Gamma_2^{(1)}$};
\end{tikzpicture}
\includegraphics[width=0.5\textwidth]{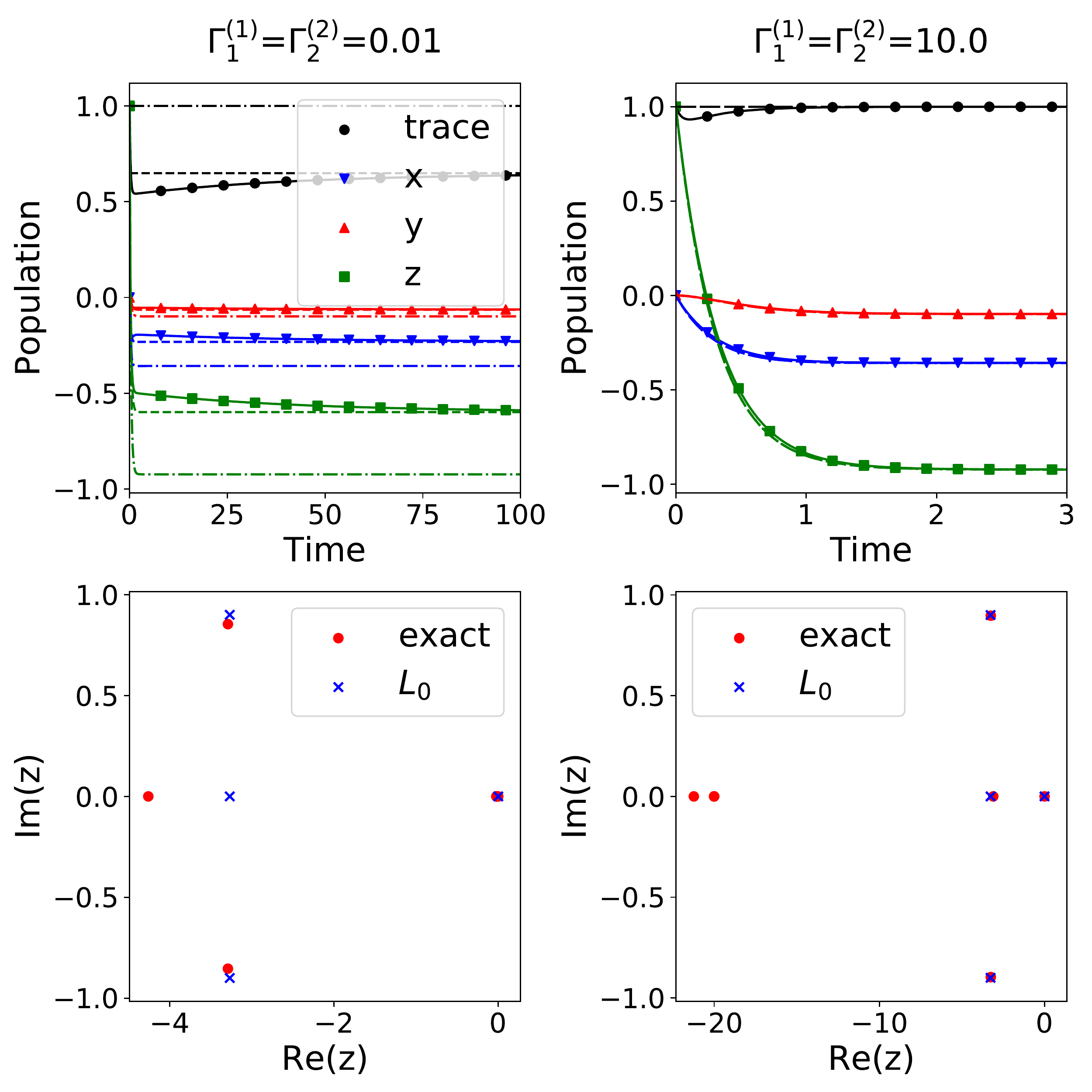}
	\caption{ \label{fig:CPT} Fano model in the zero temperature limit. The energies are given in unit of $V_1^{(1)}$ and times in units of $1/V_1^{(1)}$. Left column:  small values of the dissipation rates. Right column: high values of the dissipation rates. First row: evolution of the  expectation value of the Pauli matrices $\sigma_i$ ($i=x,y,z$) as indicated in the inset. Dash-dotted line: trace preserving evolution with $e^{L_0 t}$. Dotted line: trace rescaled evolution $\alpha e^{L_0 t}$. Second row, red circle: non linear eigenvalues of $\Leff(z)$, black cross: linear eigenvalues of $L_0$. Parameters for the simulations are 
$E_1 = 0.0$, $E_2 = 0.9$,  $V_{12} = 0.0$, $V_1^{(1)} = 1.0$, $V_2^{(1)} = 0.2$, $n=1$. All energy values are in units of $V_{1}^{1}$, so that time is in units of $\hbar/V_1^{(1)}$. The starkest difference between the exact evolution and the evolution with $U(t)=e^{L_0t}$ can be seen in the trace of the subsystem. As such this approximation sometimes fails to faithfully describe the population dynamics, which are recovered with the rescaled operator.}
\end{figure}

We calculate the time evolution with and without the correction $\alpha$ to the trace of the density matrix. 
In Fig.~\ref{fig:CPT}, we compare the exact evolution (solid line), the evolution with the effective Liouvillian
 $\rho(t) = e^{L_0 t} \rho(0)$ (dash-dotted line) and the corrected evolution with $\rho(t)=\alpha e^{L_0 t}\rho(0)$  (dashed line). For each case, we show the expectation $\text{tr}[\rho(t) \sigma_k]$ of the Pauli matrices $\sigma_k$ ($k=x,y,z$). The initial condition is $\rho(0)=\ket{g_1}\bra{g_1}$. 
For large values of the dissipation, all three evolutions coincide as expected since there is a negligible amount of population in the excited state. 
For small values of the dissipation, there is a fraction of the population that remains in the excited state so that evolution without the correction factor no longer appropriately captures the exact dynamics. 

In addition to the evolution, we show the eigenvalues  of $L_0$ and the non linear eigenvalues of $\Leff (z)$~\footnote{The non linear eigenvalues of $\Leff (z)$ are the complex numbers $\lambda$ satisfying $\left[\Leff (\lambda) -\lambda \un\right]\rho$, for some non zero $\rho$}. We see that the first eigenvalues of $L_0$ are in good agreement with those of  $\Leff (z)$.
As a consequence of the Keldysh theorem~\cite{Keldysh1951,Keldysh1971,Beyn2012} (see Appendix~\ref{app:keldysh}), the non linear eigenvalues and eigenvectors of $\Leff (z)$ completely determine the timescales of the dynamics.
 In particular the gap of $\Leff (z)$, that is the largest and non zero real part of the non linear eigenvalues of $\Leff (z)$ determine the typical time scale to reach the stationary state. We see that the gap of $\Leff (z)$ is well reproduced by the gap of $L_0$.

The correction factor $\alpha =\moy{\un+\sum_i \frac{J_i}{\Gamma}}^{-1}$ can 
also be interpreted as a detailed balance problem. To make this more 
transparent, we recognize that $\moy{L_0}=0$ so that we may write $\alpha = 
\moy{\un - \frac{N}{\Gamma}}^{-1}$ where $N=L_0-\sum_{i=1,2} J_i$ is the 
non-Hermitian Hamiltonian superoperator that describes the decay of a 
two-level system into a continuum. 
Indeed, $N = -i(\un \otimes H_D -\bar{H}_D \otimes \un)$ with 
$H_D = PHP - i \sum_{j=1,2} [F_j^{(1)}]^\dagger F_j^{(1)}$.
Therefore, the correction factor $\alpha$ can be interpreted again as the detailed balance factor arising from two sites $\sP$ and $\sQ$ equilibrating with rates $k_{\sP \to \sQ}$, corresponding to that of a non-Hermitian Hamiltonian decaying into a continuum, and $k_{\sQ \to \sP}$ corresponding to a purely incoherent transition equal to the sum of decay rates from continuum to the discrete manifold.

\subsection{Elimination of excited discrete states}
\textbf{The $\Lambda$ system}. The $\Lambda$ system is one of the most used model systems in adiabatic elimination \cite{Reiter2012}. Its usefulness lies in that it sustains most of the useful features for applications in metrology, quantum computing and thermomety, in particular in cold ion traps \cite{Radmore1982,Bergmann1998,
Boller1991,Fleischhauer2005,Vitanov2017,
Vanier1998,Sevincli2011,Kasevich1992,Morigi2000,Aspect1989,
Baron2014,Dantan2006,Schempp2010}. The Liouvillian is $L = -i(\un \otimes H -\bar{H}\otimes \un)+\sum \mathcal{D}(F_i)$ where:
\begin{equation}
\begin{split}
H_0&=E_{1}\ket{g_1}\bra{g_1}+E_2\ket{g_2}\bra{g_2}+E_3\ket{e_1}\bra{e_1} \\
H_V&=V_{12}\ket{g_1}\bra{g_2}+V_{12}^*\ket{g_2}\bra{g_1}  \\
&+V_{1}^{1}\ket{g_1}\bra{e_1}+V_{1}^{1*}\ket{e_1}\bra{g_1} \\
&+V_{2}^1\ket{g_2}\bra{e_1}+V_{2}^{1*}\ket{e_1}\bra{g_2} \\
\end{split}
\end{equation}
and the jump operators are:
\begin{equation}
\begin{split}
F_{1}^{1} &= \sqrt{\Gamma_{1}^{1}}\ketbra{g_1}{e_1} \\
F_{2}^{1} &= \sqrt{\Gamma_{2}^{1}}\ketbra{g_2}{e_1} \\
F^{1'}_{1} &= \sqrt{\Gamma_{1}^{1'}}\ketbra{e_1}{g_1} \\
F^{1'}_{2} &= \sqrt{\Gamma_{2}^{1'}}\ketbra{e_1}{g_2} \\
\end{split}
\end{equation}
The operators $F_i^{j}$ and $F_i^{j'}$ represent incoherent channels going from the excited to the ground state manifold, and from the ground state manifold to the excited state, respectively. 

\begin{figure}[h]
\centering
\begin{tikzpicture}[scale=0.7]
\draw[thick] (0,0) -- (4cm,0) node[right]{$\ket{g_1}$};
\draw[thick] (0,3cm)--(4cm,3cm) node[right]{$\ket{g_2}$};
\draw[thick]  (6cm,3cm) rectangle (7cm,3cm) node[right]{$\ket{e_1}$} ;
\draw[->,thick] (2cm,0) --(2cm,2.9cm) node[midway, right] {$V_{12}$};
\draw [->,thick] (2.5cm,0cm)--(6.25cm,3.25cm)
node[midway, above] {$V_1^{1}$};
\draw[<->,thick] (3cm,3.1cm) to[out=45,in=135] node [sloped, above] {$V_2^{1}$} (6.5cm,3.1cm);
\draw [->,thick,decorate,decoration=snake] (6.25cm,2.75cm)--(3cm,0)
node[midway,sloped ,below,] {$\Gamma_1^1$};
\draw [->,thick,decorate,decoration=snake] (6.25cm,2.75cm)--(5cm,2.75)
node[midway,sloped ,above,] {$\Gamma_2^1$};
\end{tikzpicture}
\caption{\label{fig:3LSDissip} Energy levels and transitions of a three-level system with dissipation. Hamiltonian coupling are indicated by straight arrows, dissipative processes by twisted arrows. Only decay from $\ket{e_1}$ to states $\ket{g_i}$ are represented, but the complete model include also the reverse processes, that is incoherent pumping.}
\end{figure}
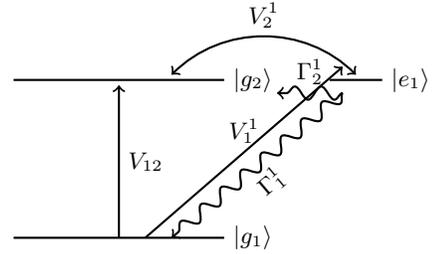

\begin{figure}[h]	\includegraphics[width=0.5\textwidth]{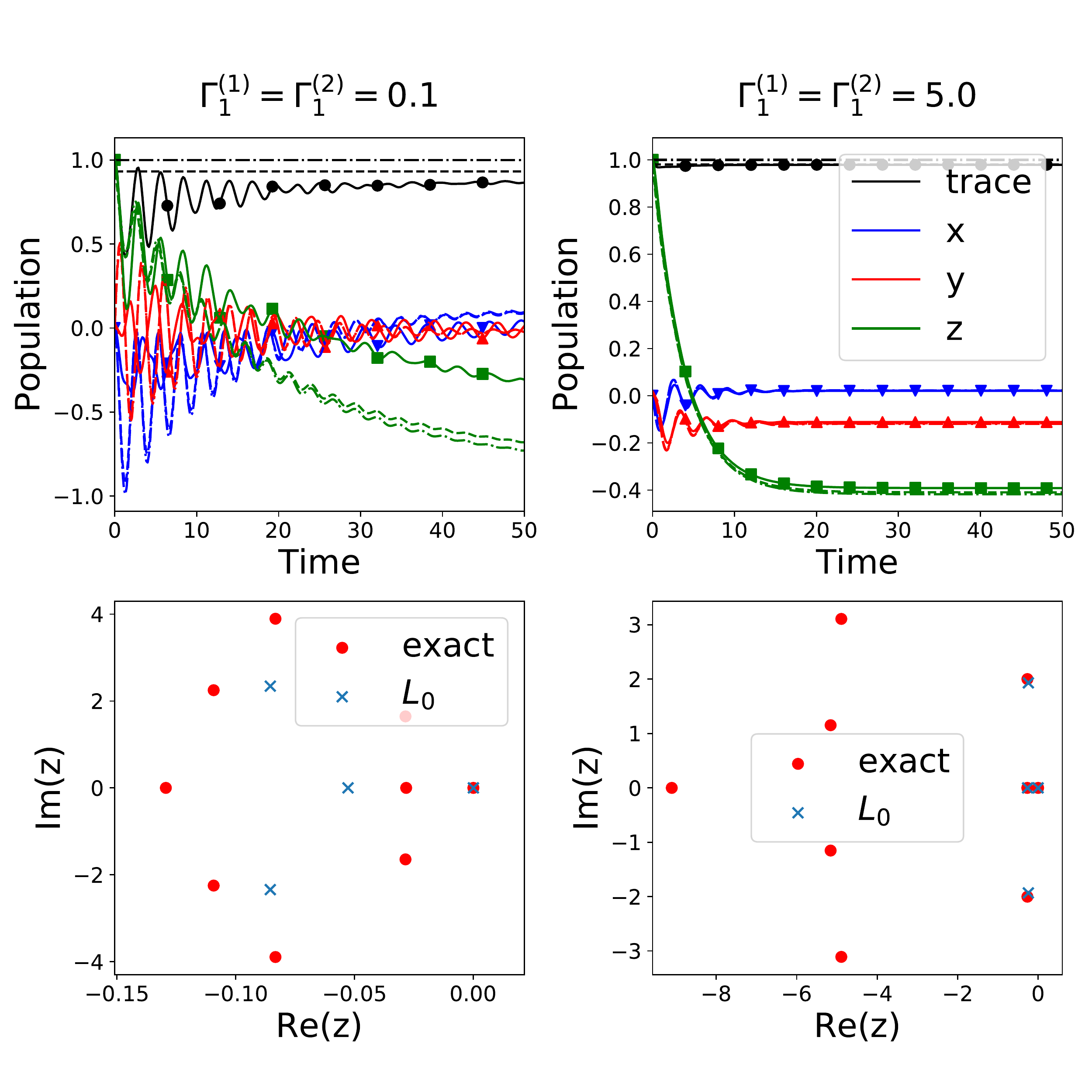}
	\caption{\label{fig:3LS_0T} Same as figure~\ref{fig:CPT}  for a $\Lambda$ system in the zero temperature limit. We show the trace of the density matrix and all three expectation values of the Pauli matrices. The parameters are 
$E_1=7$, $E_2=9$, $E_3=6$, $V_{12}=0$, $V_{1}^{1}=1.0$, $V_2^{(1)}=0.7$. All energy values are in units of $V_{1}^{1}$, so that time is in units of $\hbar/V_1^{(1)}$. The starkest difference between the exact evolution and the evolution with $U(t)=e^{L_0t}$ can be seen in the trace of the subsystem. As such this approximation sometimes fails to faithfully describe the population dynamics, which are recovered with the rescaled operator.}
\end{figure}

\begin{figure}[h]
	\includegraphics[width=0.5\textwidth]{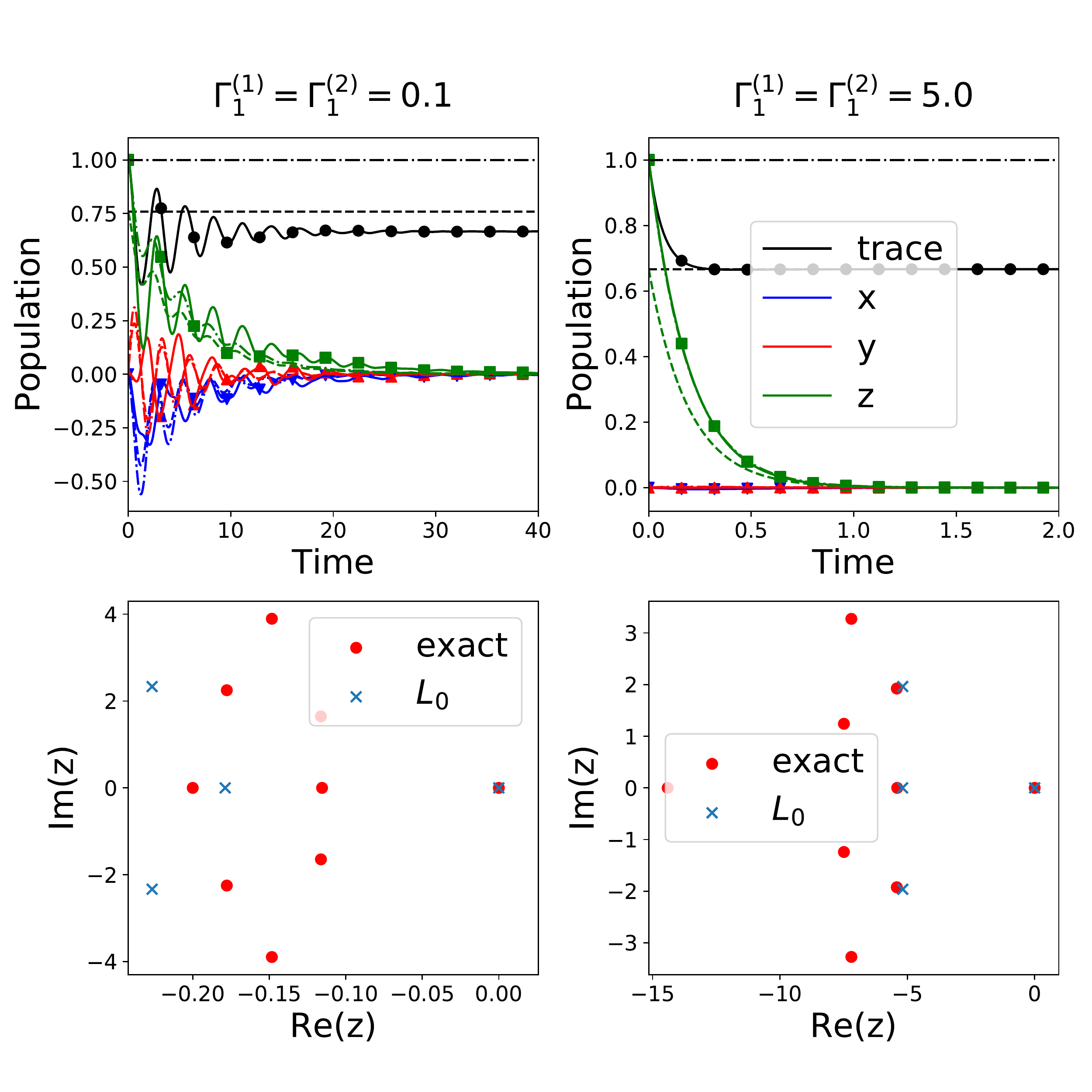}
	\caption{	\label{fig:3LS_infT} Same as figure~\ref{fig:3LS_0T} with the same parameters and in the infinite temperature limit which opens incoherent transitions from ground to excited state with identical rate as the dissipation from the excited to the ground state. We show the trace of the density matrix and all three expectation values of the Pauli matrices. The starkest difference between the exact evolution and the evolution with $U(t)=e^{L_0t}$ can be seen in the trace of the subsystem. As such this approximation sometimes fails to faithfully describe the population dynamics, which are recovered with the rescaled operator.}
\end{figure}

The effective operator $L_{\text{eff}}(z)=PLP+PLQG_0(z)QLP$ can be written in the perturbative limit up to order $\mathcal{O}((V_i^{1})^2/\Gamma)$ for $i=1,2$, and in the zero-temperature limit, as:
\begin{equation}
L_{\text{eff}}(z)=\sum_{j=1}^4 \left( \openone+\frac{M \Gamma}{z+\Gamma}  \right) \frac{h_j}{z-\xi_j}
\end{equation}
where we have used the notation $\xi_j=\{ \pm i\omega_{e_1g_1} -\Gamma/2,\pm i\omega_{e_1g_2} -\Gamma/2 \} $ and $\Gamma = \Gamma^1_1 + \Gamma_2^1$. The $h_j$ matrices are defined as follows:
\begin{equation}
\begin{split}
h_1 &= \sigma \sigma^{\dagger} \otimes S_0^{\dagger} S_0, \; \xi_1 = -i\omega_{e_1g_1} - \Gamma/2  \\
h_2 &= S_0^{\dagger} S_0 \otimes \sigma \sigma^{\dagger}, \; \xi_2 = i\omega_{e_1g_1} - \Gamma/2  \\
h_3 &= \sigma^{\dagger} \sigma \otimes S_0^{\dagger} S_0, \; \xi_3 = -i\omega_{e_1g_2} - \Gamma/2  \\
h_4 &= S_0^{\dagger} S_0 \otimes \sigma^{\dagger} \sigma, \; \xi_4 = i\omega_{e_1g_2} - \Gamma/2  \\
\end{split}
\end{equation}
where $S_0 = \sum_i V_{i}^{(1)} \ketbra{e_1}{g_i}$ and $M = -\sum_{i=1}^2\frac{M_i\Gamma_i^1}{\Gamma}$, $M_1= \sigma^{\dagger} \sigma \otimes \sigma^{\dagger} \sigma +\sigma^{\dagger} \otimes \sigma^{\dagger} $, $M_2= \sigma \sigma^{\dagger} \otimes \sigma \sigma^{\dagger} +\sigma \otimes \sigma$. The operator $\sigma$ is defined as $\sigma=\ket{g_2}\bra{g_1}$.
After some algebra we get:
\begin{equation}
\begin{split}
\Leff(z) &= L_0 + zL_1+\mathcal{O}(z^2) \\
L_0 & = \sum_{i=1}^{4} \left( \openone+M \right) \frac{h_i}{-\xi_i} \\
L_1 &= \frac{M}{\Gamma}\sum_{i=1}^{4} \frac{h_i}{\xi_i}+(\openone+M)\sum_{i=1}^{4} \frac{h_i}{-\xi_i^2} \\ 
\end{split}
\end{equation}
The form of the operators in the finite temperature limit (with incoherent 
pumping from ground to excited state) are given in the 
Appendix~\ref{app:lambdaFiniteTemp}. 
As in Fig.~\ref{fig:CPT}, in Fig.~\ref{fig:3LS_0T} and Fig.~\ref{fig:3LS_infT}, we show the evolution of the expectation of the Pauli matrices as a function of time $t$ for the same initial state. In Fig.~\ref{fig:3LS_0T} zero-temperature is considered where only dissipation from excited to discrete states takes place.
 On the contrary, in Fig.~\ref{fig:3LS_infT} the temperature is taken as infinite with equal rates for the dissipation from excited to ground and from ground to excited states. 
In the zero temperature case, there is a negligible amount of population in 
the excited state (for the perturbative calculation of $(\sQ L\sQ)^{-1}$ to 
remain valid), and both the evolution with $U(t)=e^{L_0t}$ or $U(t)=\alpha 
e^{L_0t}$ work well. As in the previous section,we notice that the gap of 
$\Leff(z)$ is very well reproduced by the one of $L_0$.

In the case of infinite temperature, the weak-field approximation is valid 
(so we can calculate the inverse of $\sQ L\sQ$ perturbatively) but there is 
a 
non-negligible population in the excited state. In this case, the correction 
$\alpha$ introduced in this article works very well in reproducing the final 
dynamics, while using a  trace preserving map does not. Writing the density 
matrix as a linear combination of Pauli matrices and the identity operator  
makes 
evident that the dynamics is well reproduced by $L_0$ (the Pauli matrices 
evolution with all operators are very close) as long as we use the correct 
normalization.

\section{Connexion between models with elimination of continuous and 
discrete 
states}

In this section we consider the connection between models where the states to be eliminated belong to a continuous set and models where  theses states are discrete. 
Although Hamiltonians with continuous spectra represent a myriad of physical systems in their own right, they can also be viewed as useful ancillary mathematical structures that make the physics behind the more complicated Hamiltonians with discrete spectrum more transparent. 
The reason for this is that Lamb shifts (or the conservative part of the level-shift operator) are absent in the case of a flat continuum (in the wideband approximation). The question we ask is: when does it matter if we describe the excited states (which we would like to eliminate) as discrete states or as approximate continua?

Intuitively, both classes of models should coincide when the population of the excited states is negligible. We will show that this happens in two cases: i) as the dissipation rate increases, the population of the excited state asymptotically vanishes and ii) at the points of coherence population trapping (CPT), the transition probability amplitudes to the excited state interfere destructively and the population of the excited state exactly vanishes \cite{Radmore1982,Bergmann1998,Boller1991,Fleischhauer2005,
Vitanov2017,Shore2017,Finkelstein-Shapiro2019} . \newline

\textbf{Coincidence for large values of the dissipation}. We calculate the limit of the effective operators $L_0$ and $L_1$, as $\Gamma/\omega_j \rightarrow \infty$. For this we recast them in terms of the smallness parameters $\delta_j = -i\omega_j/(\Gamma/2)$, for $j \in \{e_1g_1,g_1e_1,e_1g_2,g_2e_1\}$. As we take the limit of large dissipation $\lim_{\Gamma \to \infty}\delta_j = 0$ and we get for the effective operators:
 
\begin{equation}
\begin{split}
\Leff^{(\text{3LS})}(z) &= L_0^{(\text{3LS})} + zL_1^{(\text{3LS})}+\mathcal{O}(z^2), \\
L_0^{(\text{3LS})} & = \sum_j \left( \openone+M \right) \frac{h_j}{(1-\delta_j)\Gamma/2} \\
& \approx \sum_j (\openone+M)\frac{h_j}{\Gamma/2} \\
&= \frac{2}{n^{(1)} \pi \Gamma}L_0^{(\text{cont})}, \\
L_1^{(\text{3LS})} &= -\frac{M}{\Gamma}\sum_j \frac{h_j}{(1-\delta_j)\Gamma/2}\\
&+(\openone+M)\sum_j \frac{h_j}{-(1-\delta_j)^2(\Gamma/2)^2} \\ 
& \approx \frac{2}{n^{(1)} \pi \Gamma} L_1^{(\text{cont})} - \frac{2}{n^{(1)} \pi \Gamma}L_0^{\text{(cont)}} \\ 
\end{split}
\end{equation}
where the labels (3LS) and (cont) mean 3-level system and continuum models respectively.
We find that in the limit of large dissipation, the continuum and discrete effective operators for $L_0$ are the same as long as we set the density of states in the continuum model as $n=2/(\pi \Gamma)$, while they differ for $L_1$. 
We can understand this convergence of operators as follows. 
The level-shift operator for the discrete excited states consists of a real part related to the dissipation and an imaginary part related to the Lamb shift. That of a flat continuum only has the real dissipative part. As the dissipation rate increases, the Lamb shift part of the operator becomes negligibly small and a discrete excited state becomes analogous to a continuum manifold as far as the evolution of the ground states are involved.

\begin{figure}
\includegraphics[width = 0.5\textwidth]{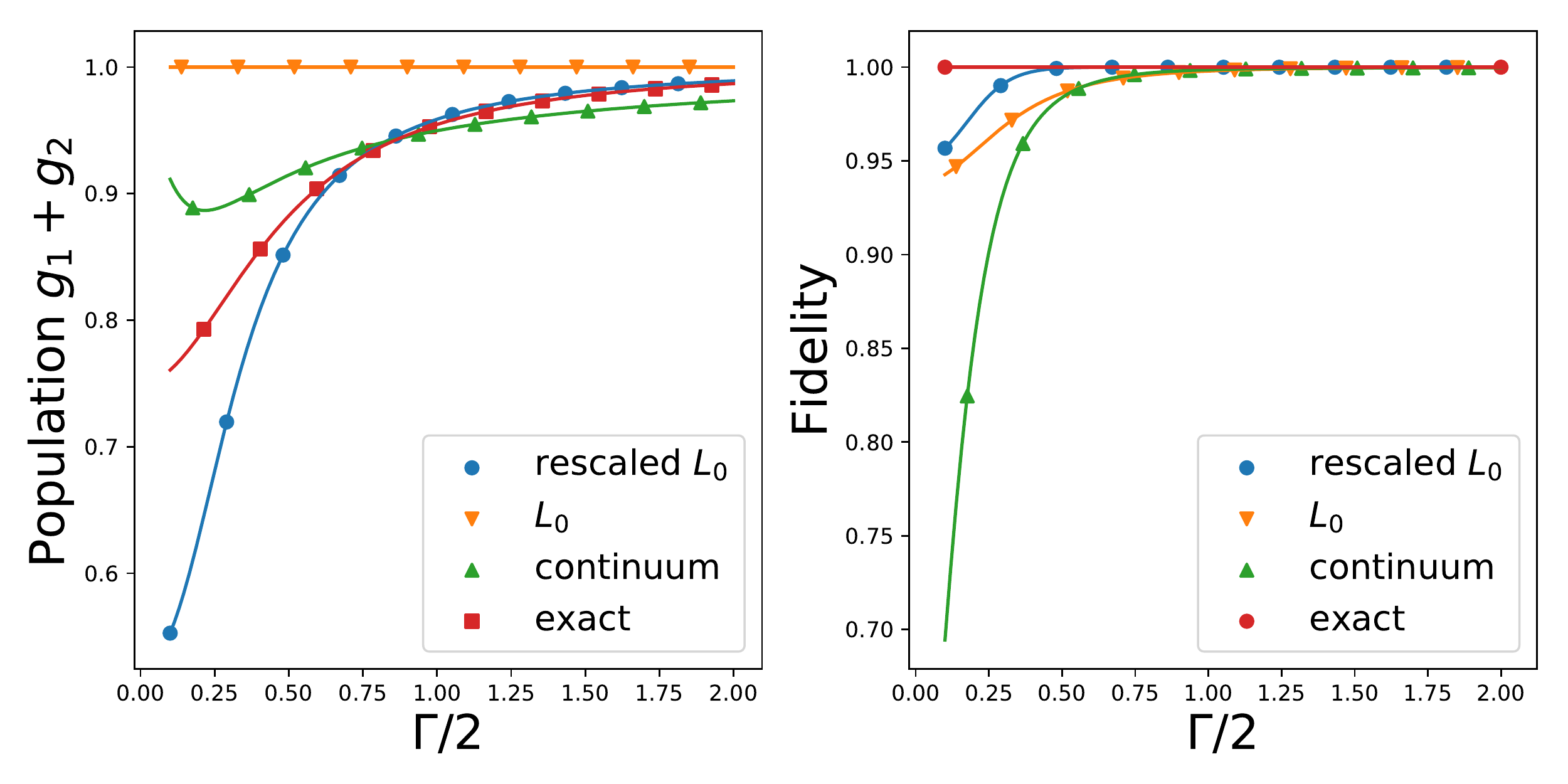}
\caption{Ground state population and fidelity of the steady-state density 
matrix evolved according to the exact and several approximate operators as a 
function of the dissipation rate. The parameters of the calculation are 
$E_1=0.5$, $E_2=-0.1$, $E_3=0.01$, $V_{12}=0$, $V_1^{(1)}=0.2$, 
$V_{2}^{(1)}=0.3$. }
\label{fig:fidelity}
\end{figure}

In Fig.~\ref{fig:fidelity}, we show the convergence of these models towards the exact solution of a $\Lambda$ system. For this we plot the steady-state population in the ground states and the steady-state fidelity as a function of dissipation rate from excited states to ground states. We rescale the  fidelity $F=\text{Tr}\left(\sqrt{ \sqrt{\rho_{\text{exact}}} \rho \sqrt{\rho_{\text{exact}}} }\right)^2/(\text{Tr}(\rho)\text{Tr}(\rho_{\text{exact}}))$ by a factor $1-(\text{Tr}(\rho)-\text{Tr}(\rho_{\text{exact}}))^2$ which penalizes evolution operators that do not have the correct asymptotic trace. 
We clearly see that the rescaled steady-state for a three-level system performs best, and that the rescaled steady-state for an equivalent continuum and the unscaled steady-state for the three-level system approach the correct solution for similar values of the dissipation rate. \newline

\textbf{Coincidence at the coherence population trapping points}. It can be shown that as long as we do not have dissipation within the ground state manifold, there will be points of coherence population trapping as long as the following conditions are fulfilled \cite{Finkelstein-Shapiro2019}: 

\begin{equation}
\begin{split}
[PHP,\rho] &= 0 \\
QHP = 0
\end{split}
\end{equation}

Remarkably, this condition is independent of the value of the dissipation rate, so that we are free to choose an arbitrarily large value and still retain the property of CPT where the population is restricted to the ground state manifold. Accordingly, it follows from the previous paragraph that if we scale $n^{(1)}=2/(\pi\Gamma)$ then the effective operators will be the same. 
It also follows that since $\alpha=1$, then $\left \langle L_1 \right \rangle=0$.

\begin{figure}
\includegraphics[width = 0.5\textwidth]{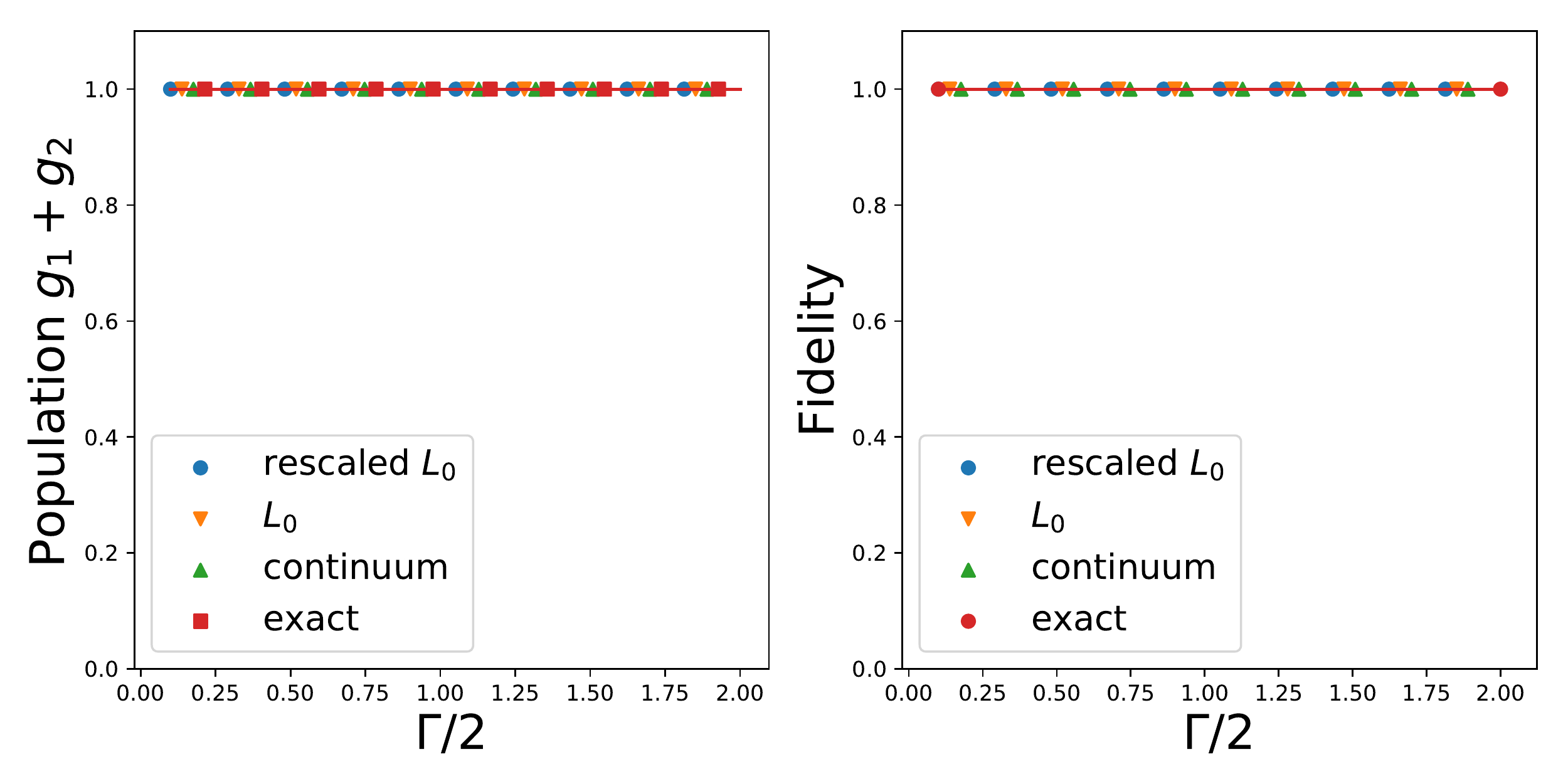}
\caption{Ground state population and fidelity of the steady-state density 
matrix evolved according to the exact and several approximate operators as a 
function of the dissipation rate, at the CPT condition. All traces overlap.}
\label{fig:fidelity_CPT}
\end{figure}

\begin{figure}
\includegraphics[width = 0.5\textwidth]{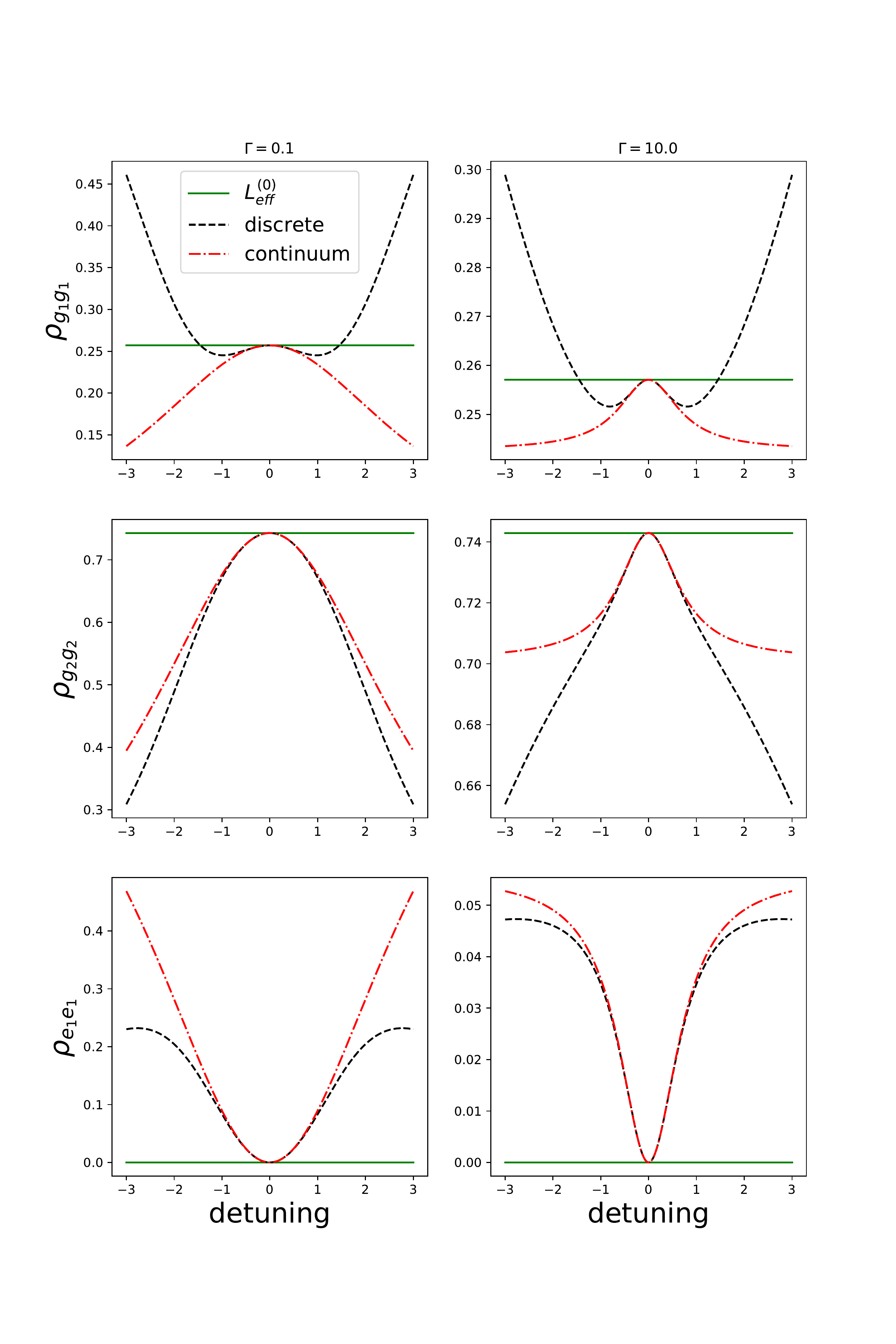}
\caption{Steady-state population in the ground state manifold $\rho_{g_1g_1}$ and $\rho_{g_2g_2}$, and in the excited state manifold $\rho_{e_1e_1}$, with a three-level system, a Fano model and an effective Liouvillian evolution with $L_0$ for the three-level system. Parameters are $V_1^{(1)}=1.7$, $V_2^{(1)}=1.0$. The detuning between $g_2$ and $e_1$ is set to zero while the detuning between $g_1$ and $e_1$ is varied. All energies are given in units of $V_2^{(1)}$.}
\label{fig:steady-state_CPT}
\end{figure}

We illustrate the effect of coherence population trapping points on our models in Figures \ref{fig:fidelity_CPT} and  \ref{fig:steady-state_CPT}. By plotting the ground state population and fidelity of the three models at the CPT condition we see that all four models coincide (Fig. \ref{fig:fidelity_CPT}). To further stress the equivalence of the models around CPT, we plot the steady-state of a three-level system, of a Fano model and of the effective Liouvillian $L_0$ for a three-level system, as a function of the detuning of the ground states $E_{g_1}-E_{g_2}$ (Fig. \ref{fig:steady-state_CPT}).
We observe the CPT point at zero detuning where all three models coincide. The unscaled $L_0$ only agrees at the CPT condition since it preserves the population in the ground state manifold while both the continuum and the exact $\Lambda$ system agree around a neighborhood of the CPT point. \newline

We have shown that replacing discrete excited states by continua corresponds to taking the limit of large dissipation, or alternatively finding the CPT points. This is important since the effective operator with a continuum is much more straightforward to calculate exactly than that of a discrete level. Thus calculations that fulfill these conditions, if carried out using these simplified operators, can be more easily solved analytically.

\section{Conclusion}

We have derived expressions for the adiabatic elimination of a fast manifold in frequency space. This has allowed us to correct for particle density loss to the fast manifold and rescale the evolution operator. 
We have illustrated this with examples spanning discrete and excited state continua which show the advantages of the correction factor as well as its physical meaning. 
We have provided an equivalence between the discrete and continuum models at the CPT condition and in the limit of large dissipation, giving insight into commonly used adiabatic elimination approaches. \newline

\textbf{Acknowledgements.} D.F.S.  acknowledges  support from the European Union through the Marie Sklodowska-Curie Grant Agreement No.  590 702694. \newline

\appendix

\section{Operator vs. superoperator notation}
\label{app:superoper_notation}

For an $N$-level system, the underlying Hilbert  space $\mathcal{H}$ is of 
dimension $N$ and the states of the quantum systems are described by 
positive operators acting on $\mathcal{H}$ that can be represented by  
$N\times N$ density matrices. The superoperators as the Lindblad operator 
$L$ or its resolvent $G$ are linear operators acting on operators themselves 
acting on $\mathcal{H}$.

To describe an open quantum system we need to know the evolution of the density matrix using Lindblad equation, which in Hilbert $\mathcal{H}$ space is written as:
\begin{equation}
\dot{\rho}(t) = L\rho = -i[H,\rho(t)]+\sum_i \left( F_i\rho F_i 
^{\dagger}-\frac{1}{2} \{F_i^{\dagger}F_i, \rho(t)   \} \right)
\end{equation}
or as:
\begin{equation}
\rho(t) = e^{Lt}\rho(0) =  \frac{1}{2\pi i} \int \ud z\frac{e^{zt}}{z-L} 
\rho(0) 
\end{equation}
A disadvantage of this form is that neither the exponential map nor the 
resolvent can be straightforwardly expressed or calculated numerically. It
is therefore convenient to represent the density matrix as a vector 
$\vec{\rho}$ with $N^2$ components, obtained from  the column-stretched form 
of the $N\times N$ density matrix. 
This representation is obtained by considering the density matrix as an 
element $\vec{\rho}$ of the Hilbert space $\mathcal{H}\otimes 
\mathcal{H}$~\cite{Havel2003}. 
In that way, superoperators are linear operators acting on 
$\mathcal{H}\otimes \mathcal{H}$ and they can be  represented by 
$N^2\times N^2$ matrices. The linear superoperator acting on $\vec{\rho}$,  
built from 2 arbitrary 
operators $S_n$ and $S_m$ on $\sH$ and  acting on $\rho$ as $S_n\rho 
S_m^{\dagger}$ is given by
the mapping    $S_n\rho S_m^{\dagger} \to 
(\bar{S}_m \otimes S_n \vec{\rho})$. 
With 
the help of this mapping, the Lindblad operator operating on the vector form 
of 
the  density matrix as $\frac{\ud}{\ud t}\vec{\rho}(t) = L\vec{\rho}(0)$ is
:
\begin{equation}
\label{eq:Loperator}
\begin{split}
L &= -i (\un \otimes H - \bar{H} \otimes \un) \\
& + \sum_i \left[ \bar{F}_i \otimes F_i  - \frac{1}{2}\left(\un \otimes 
F_i^{\dagger} F_i + (F_i^{\dagger} F_i)^T \otimes \un \right) \right]
\end{split}
\end{equation}
\section{Perturbative inversion of $\sQ L \sQ$}
\label{app:pertubInversion}
We consider a generic system, with a Hamiltonien $H$ written as~:
$H = H_0 + V$, where $H_0 = PHP + QHQ$ and $V = PHQ + QHP$. Let
$\ket{i;p}$ ($\ket{j,q}$) the eigenstates of $PHP$ ($QHQ$), with 
$i=1,2,\cdots N_p$ ($j=1,2,\cdots N_q$). 
For the dissipation processes, we consider relaxation from the fast subspace 
$\ran{Q}$ to the slow subspace $\ran{P}$, described by jump operators 
$F_{ij} = \sqrt{\Gamma_{ij}}\ketbra{i;p}{j;q}$, relaxation from from the 
slow subspace to the fast subspace described by jump operators $J_{ji} = 
\sqrt{\gamma_{ij}}\ketbra{j;q}{i;p}$ and finally
 we also consider relaxation inside 
$\ran{P}$, described by jump operators $N_m$ which we don't specify as they 
don't intervene in $QLQ$. We neglect all the dissipation processes between 
states belonging to $\ran{Q}$.  

It is convenient to define a non hermitian Hamiltonian operator $K= K_0 + V$ 
where
\[
K_0 = H_0 - \frac{\imath}{2}\left( \sum_{ij} F_{ij}^\dagger F_{ij} + 
\sum_{ij} 
J_{ji}^\dagger J_{ji} + 
\sum_n{N_m^\dagger}N_m \right),
\]
has a diagonal matrix representation in the basis $\left\{\ket{i;p}, 
\ket{j,q}\right\}$. 
we can the rewrite the Lindblad operator $L$ as (see 
Eq.~\eqref{eq:Loperator})~:
\begin{align*}
L = -\imath \left(\un \otimes K - \bar{K} \otimes \un\right) 
&+ \sum_{ij} \left(  \bar{F}_{ij} \otimes F_{ij} +    \bar{J}_{ji} \otimes 
J_{ji}\right) \\
&+  \sum_{n} \bar{N}_{n} \otimes N_{n} 
\end{align*}
Using the expression of $\sQ$ given by Eq.~\eqref{eq:defPQ}, we notice that
$\bar{F}_{ij} \otimes F_{ij} = \sP \bar{F}_{ij} \otimes F_{ij} \sQ$, 
$\bar{J}_{ji} \otimes J_{ji} = \sQ \bar{J}_{ji} \otimes J_{ji} \sP $ and
$\bar{N}_{n} \otimes N_{n} = \sP \bar{N}_{n} \otimes N_{n} \sP$.

Therefore $\sQ L \sQ$ can be written as:
\[
\sQ L \sQ  = L_D + W
\]
Where $L_D = -\imath \sQ\left(\un \otimes K_0 - \bar{K}_0 \otimes 
\un\right)\sQ$ has a diagonal matrix representation in the basis 
$\left\{\ket{i;p}\otimes
\ket{j,q}\right\}$ and $W = -\imath \sQ\left(\un \otimes V - V \otimes 
\un\right)\sQ$ has a non diagonal matrix representation in the same basis.

The non diagonal part of $\sQ L \sQ$ depends only upon the Hamiltonian 
coupling $V$ which can be considered as a small perturbation with respect to 
the diagonal part when the relaxation of the fast space $\ran{Q}$ is fast 
($\Gamma_{ij}\gg V{ij}$). 
 
\section{General case of $N_g$ ground states coupled to $N_e$ continua \label{app:general_continua}}
We provide here the general expressions to calculate the effective 
Liouvillian for $N_g$ discrete ground states coupled to $N_e$ continua, from 
which the more specific examples detailed in the main text can be derived. 
The complete Liouvillian for such a system is:

\begin{align}
&H=H_0+H_V+H_{V_p} \\
&H_0=E_i\ket{i}\bra{i}+\sum_a\int dk_a E_{k_a}\ket{k_a}\bra{k_a} \nonumber \\
&H_V=\sum_a \int dk_a \big[ V_{i}^{(a)}\ket{i}\bra{k_a}+(V_{i}^{(a)})^*\ket{k_a}\bra{i} \big] \nonumber \\
&H_{V_p}=\sum_{i,j} \big[ V_{ij}\ket{i}\bra{j}+V_{ij}^*\ket{j}\bra{i} \big] \nonumber \\
\label{eq:Hamiltonian}
\end{align}
With the dissipative part the Liouvillian is:
\begin{equation}
L= -i(1\otimes H - \bar{H}\otimes 1)+\sum_{i,a}\mathcal{D}(F_i^{(a)})
\end{equation}
\begin{equation}
F_i^{(a)} = \sqrt{\Gamma_{i}^{(a)}}\ketbra{g_i}{k_a}
\end{equation}

The effective operators are obtained in a similar calculation as we have done previously \cite{Finkelstein2017}
 but keeping the $z-$dependence of the operators. Briefly, we define the projection operators for the continuous part ($\sQ$) and the discrete part ($\sP$). The effective Liouvillian (see Equation~\eqref{eq:LeffDef}) hinges on the resolvent operator in $\sQ$. This operator can be expanded in a Lippman-Schwinger series that is exactly resummed for the wideband approximation, where the parameters of the continuum do not depend on the continuum energy. 
 We obtain:
 \begin{equation}
 \begin{split}
 \Leff(z) &= -i(1\otimes H - \bar{H} \otimes 1) +\sum_{i}^{N_g}\sum_a^{N_e} \mathcal{D}({F}_{\text{eff},i}^{(a)}) \\
& +\sum_{i}^{N_g}\sum_a^{N_e} f^{(a)}(z) \bar{F}_{\text{eff},i}^{(a)} \otimes {F}_{\text{eff},i}^{(a)}
 \end{split}
 \end{equation}
 where ${F}_{\text{eff},i}^{(a)} = \sum_j\sqrt{ \frac{\Gamma_i^{(a)}}{\sum_l \Gamma_l^{(a)}} n^{(a)} \pi}V_j^{(a)}\ketbra{i}{j}$ and $f^{(a)}(z) = - \frac{z}{z+\sum_l \Gamma_l^{(a)}}$ clearly vanishes when $z=0$. We recognize that the nonlinear operator can be written as a $z$-independent part in Lindblad form and a $z$ dependent part which involves only the quantum jump that restores population to the ground state. 
From the above expressions the specific cases in the Examples section can be straightforwardly derived. 

\section{$\Leff(z)$ for a $\Lambda$ system at finite temperature}
\label{app:lambdaFiniteTemp}
We give the general expression for the effective operator of a $\Lambda$ system with incoherent transitions from the ground-state manifold to the excited states. The generalization of the operator presented in the main text is:
\begin{equation}
L_{\text{eff}}(z)= \left( \openone+\frac{M \Gamma}{z+\Gamma}  \right) \sum_{j=1}^4 \frac{h_j}{z-\xi_j} \left( \openone+\frac{M' \Gamma'}{z+\Gamma}  \right) + \frac{1}{2}\frac{MM'\Gamma \Gamma'}{z+\Gamma}
\end{equation}
where we have used the notation $\xi_j=\{ \pm i\omega_{e_1g_1} -\Gamma/2-\Gamma'^1_1/2,\pm i\omega_{e_1g_2} -\Gamma/2-\Gamma'^1_2/2 \} $ and $\Gamma = \Gamma^1_1 + \Gamma_2^1$. The $h_j$ matrices are defined as follows:
\begin{equation}
\begin{split}
h_1 &= \sigma \sigma^{\dagger} \otimes S_0^{\dagger} S_0, \; \xi_1 = -i\omega_{e_1g_1} - \Gamma/2 -\Gamma'^1_1/2 \\
h_2 &= S_0^{\dagger} S_0 \otimes \sigma \sigma^{\dagger}, \; \xi_2 = i\omega_{e_1g_1} - \Gamma/2 -\Gamma'^1_1/2 \\
h_3 &= \sigma^{\dagger} \sigma \otimes S_0^{\dagger} S_0, \; \xi_3 = -i\omega_{e_1g_2} - \Gamma/2 -\Gamma'^1_2/2 \\
h_4 &= S_0^{\dagger} S_0 \otimes \sigma^{\dagger} \sigma, \; \xi_4 = i\omega_{e_1g_2} - \Gamma/2 -\Gamma'^1_2/2 \\
\end{split}
\end{equation}
where $S_0 = \sum_i V_{i}^{(1)} \ketbra{e_1}{g_i}$, $M = -\sum_{i=1}^2\frac{M_i\Gamma_i^1}{\Gamma}$, $M_1= \sigma^{\dagger} \sigma \otimes \sigma^{\dagger} \sigma +\sigma^{\dagger} \otimes \sigma^{\dagger} $, $M_2= \sigma \sigma^{\dagger} \otimes \sigma \sigma^{\dagger} +\sigma \otimes \sigma$ and 
$M' = -\sum_{i=1}^2\frac{M'_i \Gamma_i^{'1}}{\Gamma'}$, 
$\Gamma'=\Gamma_1^{'1}+ \Gamma_2^{'1}$, $M'_1 = \sigma^{\dagger}\sigma \otimes \sigma^{\dagger}\sigma + \sigma \otimes \sigma$, $M'_2 = \sigma\sigma^{\dagger} \otimes \sigma \sigma^{\dagger} + \sigma^{\dagger} \otimes \sigma^{\dagger}$ 
. The operator $\sigma$ is defined as $\sigma=\ket{g_2}\bra{g_1}$.

\section{$L_0$ generator of a trace preserving dynamics}
\label{app:L0TracePreserv}
Let us recall the expression for $L_0$:
\begin{equation}
L_0=\sP L\sP-\sP L\sQ\left(\sQ L\sQ\right)^{-1}\sQ L\sP.
\end{equation}
We can rewrite this equation as:
\beq
L_0 =L\mathcal{A}-\mathcal{G}
\eeq
where we have defined the operators $\mathcal{A}$ and $\mathcal{G}$ as follows:
\begin{eqnarray}
\mathcal{A}=&\sP-\left(\sQ L\sQ\right)^{-1}\sQ L\sP,\\
\mathcal{G}=&\left(\un-\sQ L\sQ\left(\sQ L\sQ\right)^{-1}\right)\sQ L\sP=\sQ' \sQ L\sP.
\end{eqnarray}
In all above equations, $A^{-1}$ signifies the Moore-Penrose inverse of $A$~\cite{Penrose1955} which coincides with the matrix inverse when $A$ is invertible. Finally, we have defined $\sQ'=\un-\sQ L\sQ\left(\sQ L\sQ\right)^{-1}$ which is an orthogonal projector~\cite{golub13} to $\ker{\left[\left(\sQ L \sQ\right)^{\dagger}\right]}=\text{ran}\left[\sQ L \sQ\right]^\perp$, where $^\perp$
stands for orthogonal complement.\\

In order for $L_0$ to be a generator of a trace preserving map, the maximally mixed state $[\rho] = \frac{1}{N}\un_{\mathcal{H}}$, 
must be a left eigenvector for $L_0$ with eigenvalue $0$, where $N$ is the dimension of $\sH$.
In vector form, we can associate to $\vec{\rho}$,
 the maximally entangled state $\ket{\un_{\mathcal{H}}} \in \sH\otimes\sH$. Therefore, the trace preserving condition can be written as:
\begin{equation}
L_0^{\dagger}\ket{\un_{\mathcal{H}}}=\mathcal{G}^\dagger\ket{\un_{\mathcal{H}}}=0,
\end{equation}
where we have used the fact that $L$ is a Lindblad operator, hence generating a trace preserving dynamics. If we define the set $\mathcal{X}$ as
\begin{equation}
\mathcal{X}=\text{ran}\left[\sQ L \sQ\right]^\perp\cap\text{ran}\left[\sQ L \sP\right],
\end{equation}
then a sufficient condition for $L_0$ to generate a trace preserving dynamics is that $\ket{\un_{\mathcal{H}}}$ is orthogonal to the set $\mathcal{X}$. Equivalently, since $\mathcal{X} \subseteq \text{ran}\left[\sQ\right]$, we can write this condition as
\begin{equation}
\ket{\Psi_{\sQ}}\perp \mathcal{X}
\end{equation} 
where we have defined the state $\ket{\Psi_{\sQ}}=\sum_{i_q}\ket{i_q}\otimes\ket{i_q}$, with the index $i_q$ enumerating the left eigenvectors of $Q$ corresponding to eigenvalue $1$.

In all examples considered in this article, we had $\text{ran}\left[\sQ L \sP\right]\subseteq \text{ran}\left[\sQ L \sQ\right] = \ran{Q}$ implying that $\mathcal{X}$ is an empty set and, therefore, the fulfillment of the above condition.
\section{Keldysh theorem}
\label{app:keldysh}
For the sake of completeness, we recall here the Keldysh theorem. We consider only the case where the non linear eigenvalues are simple. 
This section is based on the material of Ref.~\cite{Beyn2012}.
To connect our notation with the usual statement of the theorem, we define  $T(z)$ such that $T(z)=z\un -\Leff(z)$, therefore
$\sP G(z)\sP = \left[T(z)\right]^{-1}$.

First we recall the definition of a nonlinear eigenvalue $\lambda$ of $T(z)$: $\lambda$ is an eigenvalue of $T(z)$ if $T(\lambda) v=0$ for some nonzero vector $v$. the vector $v$ is the right eigenvector of $T$. The eigenvalue is called simple if in addition:
\[
\ker[T(\lambda)] = \text{span}\{v\};\quad v\neq 0;\quad T'(\lambda) \notin \text{ran}[T(\lambda)].
\]
In this case the adjoint $T^{\dagger}$ of $T$ satisfies:
\[
\ker[T^{\dagger}(\lambda)]= \text{span}\{w\}
\]
for some non zero vector $w$, and furthermore,
$w^{\dagger}T'(\lambda)v \neq 0$. Without loss of generality we can choose
\beq
w^{\dagger}T'(\lambda)v = 1
\eeq
where $T'(\lambda)$ is the value of the derivative of $T(z)$ with respect to $z$, taken at $z=\lambda$.

The Keldysh theorem states that: let $D$ be a compact subset that contains only simple eigenvalues 
$\lambda _n, \quad n=1,\cdots, N$, with right and left eigenvectors $v_n$ and $w_n$, respectively, then there 
is a neighborhood $U$ of $D$ and a holomorphic function $R(z)$
such that
\beq
T(z)^{-1} = \sum_{n=1}^{N} \frac{1}{z-\lambda_n}v_n w_n^{\dagger} + R(z).
\label{eq:keldyshTheo}
\eeq

Now, if we assume that all the eigenvalues of $T$ are simple, then we can use Eq.~\eqref{eq:keldyshTheo} to calculate $\sP \rho(t)$, performing the integration of Eq.\eqref{eq:P-evolution}, and we obtain:
\beq
\label{eq:keldyshEvolution}
\sP \rho(t) = \sum_{n=1}^{N} e^{\lambda_n t} v_n w_n^{\dagger}\rho(0).
\eeq.

\section{Correction to the steady-state trace}
\label{app:tracecorrection}
We know that the steady-state $\rho_f$  of the system will be in the kernel of $L_0$ that is $\rho_f = \alpha \bar{\rho}$ where $L_0\bar{\rho} = 0$ and $\tr{\bar{\rho}} = 1$.
We are only left with determining the constant $\alpha$. This can be done from the final value theorem: 
\beq
\rho_f = \lim_{z\rightarrow 0} z\sP G(z) \sP\rho(0) = 
 \lim_{z\rightarrow 0} z \left[z-\Leff(z)\right]^{-1} \rho(0)
\eeq
We expand $\Leff(z)$ as $\Leff(z) = L_0+zL_1+z^2L_2+\cdots$ and get:
\begin{equation}
\begin{split}
\rho_f &= \lim_{z\to 0} z[z-\Leff(z)]^{-1} \rho(0) \\
&= \lim_{z\to 0} \frac{z}{z(\un-L_1)-L_0+\mathcal{O}(z^2)} \rho(0).
\end{split}
\end{equation}
Multiplying by $(\un -L_1)$, and taking the limit, we obtain:
\[
[\un - L_1]\rho_f  = [\un - L_1]\alpha \bar{\rho} = \lim_{z\to 0}[\un - \frac{1}{z}L_0(\un -L_1)^{-1}]^{-1}\rho(0),
\]
taking the trace of both side, we obtain
\begin{equation}
\alpha = \frac{1}{\tr{(\un-L_1)\bar{\rho}}}
\label{eq:alpha}
\end{equation}
where we have used the fact that the dynamics generated by $L_0$ is trace preserving implying that $\tr{L_0\rho} = 0$ for all operator $\rho$  and where we have considered that $\tr{\rho(0)}=1$.

The same result can be obtained using the Keldysh theorem. Indeed, 
taking the limit $t\rightarrow \infty$ of Eq.~\eqref{eq:keldyshEvolution}, we get $
\rho_f = \bar{\rho} w_0^\dagger \rho(0) = \alpha \bar{\rho}
$
and thus $\alpha = w_0^\dagger \rho(0)$,
which in matrix form means
\[
\alpha = \tr{w_0^\dagger \rho(0)},
\]
where $w_0$ is such that $L_0^\dagger w_0 = 0$, and 
$w_0^\dagger T'(0)\bar{\rho}= 1$ that is, 

\beq
w_0^\dagger( \un - L_1)\bar{\rho}= 1.
\label{eq:normKeldysh}
\eeq
But in matrix form, $[w_0]$ is proportional to the identity,$[w_0]= \beta \un_{\sH}$. This is a consequence of the trace preserving dynamics induced by $L_0$. 
Therefore, Eq.~\eqref{eq:normKeldysh} gives $\beta^* \tr{( \un - L_1)\bar{\rho}} = 1$, and
$\alpha = \beta^*\tr{\rho(0)}$. Considering that $\tr{\rho(0)}=1$, we
obtain the same result as in Eq.~\eqref{eq:alpha}.

\bibliography{Fano,dark-states,adiabElimin}

\end{document}